\documentclass[apjl,twocolumn]{openjournal}

\usepackage{natbib} 
\usepackage[dvipsnames]{xcolor}
\usepackage{aas_macros} 
\usepackage{amssymb}
\usepackage{amsmath}
\usepackage{threeparttable}
\usepackage{multirow}
\usepackage{comment}
\usepackage[title]{appendix}
\usepackage{hyperref}	
\usepackage{orcidlink}
\hypersetup{colorlinks=true,linkcolor=blue,citecolor=blue,filecolor=blue,urlcolor=blue}
\hypersetup{
    colorlinks=true,       
    linkcolor=cyan,          
    citecolor=cyan,         
    urlcolor=cyan        
}
\usepackage[caption=false]{subfig}
\usepackage{soul}

\newcommand{\modelname}{\textsc{Phot-Gal }}
\newcommand{\modelnamec}{\textsc{Phot-Gal}}

\graphicspath{ {./figures/} }

\usepackage{relsize}
\newcommand{\authsize}{\fontsize{10}{14}\selectfont} 

\usepackage{etoolbox}

\makeatletter
\patchcmd{\normalsize}{\@setfontsize\normalsize\@xpt{10.56}}
  {\@setfontsize\normalsize\@xpt{10.56}\baselineskip=1.1\baselineskip}
  {}{}
\makeatother

\makeatletter
\patchcmd{\@maketitle}
  {\small}  
  {\large}  
  {}{}
\makeatother

\begin{document}
\title{Reimagining SED Fitting with Cosmological Galaxy Simulations and Machine Learning}

\author{{\authsize Dhruv T. Zimmerman} \orcidlink{0009-0008-7017-5742}$^{1}$}
\author{{\authsize Desika Narayanan} \orcidlink{0000-0002-7064-4309}$^{1,2}$}


\affiliation{$^{1}$Department of Astronomy, University of Florida, 211 Bryant Space Sciences Center, Gainesville, FL 32611 USA}
\affiliation{$^{2}$Cosmic Dawn Center (DAWN), Niels Bohr Institute, University of Copenhagen, Jagtvej 128, København N, DK-2200, Denmark}

\thanks{$^*$e-mail: \href{mailto}{d.zimmerman@ufl.edu}}



\begin{abstract}

SED fitting is the most common technique to recover galaxy physical properties from observed photometry. However, SED fitting requires many assumptions that essentially collapse a galaxy from a three-dimensional spatially varying object with complex structure into a scalar point. Moreover, modern inference techniques are computationally intensive, which presents a unique challenge in the era of extremely large datasets. We present \modelnamec, a new galaxy SED modeling tool that solves the inverse problem of SED fitting by training a machine learning model on photometry generated from 3D radiative transfer of simulated galaxies with a wide range of implemented physics. \modelname is designed to accept an arbitrary amount of input photometry by utilizing a $K$-nearest neighbors imputation strategy. Our fiducial model predicts redshift, stellar mass, dust mass, and star formation rate with uncertainties based on the provided input photometry. We evaluate the performance of \modelname relative to the commonly-used SED fitting tool \textsc{prospector} in successfully recovering each of these properties with several metrics for the inferred values and uncertainties and find that it outperforms the accuracy of standard SED fitting software on the testing set. However, with fewer photometric constraints, \modelname is more likely to have output uncertainties that do not reflect the offset from the ground truth. We dissect the components of \modelname to find reasonable physical justifications for the photometry it relies on most, understand how each step in its workflow contributes to the eventual output posterior, and evaluate its ability to generalize to novel data.

\end{abstract}

\section{Introduction}\label{sec:intro}

Spectral Energy Distribution (SED) modeling is the most common technique for deriving the physical properties from the photometry and spectra of galaxies near and far. Originally introduced by \cite{tinsley_evolution_1968}, \cite{spinrad_stellar_1971}, and \cite{faber_quadratic_1972}, this approach involves modeling stellar population emission as it changes over a given star formation history and accounting for dust attenuation effects that vary with wavelength \citep{walcher_fitting_2011,conroy_modeling_2013,salim_dust_2020}. In practice, these are varied until the resulting mock spectral shape fits observational constraints, at which point the physical properties of the galaxy are assumed to be known from the model.  Some common SED fitting packages used in the literature are \textsc{prospector}, \textsc{bagpipes}, \textsc{cigale}, and \textsc{magphys} \citep{leja_deriving_2017,carnall_inferring_2018,boquien_cigale_2019,da_cunha_magphys_2012}.

While our community's knowledge of cosmological galaxy evolution rests in large part on the collective results of SED fitting from decades of observations, the inference of key galaxy physical properties rely on numerous assumptions, including the stellar initial mass function (IMF), stellar isochrones, spectral libraries, star formation history (SFH) shape, and the dust attenuation law. These assumptions can systematically affect the inferred galactic properties. For example, it is common in many SED fitting packages to assume a parametric functional shape to the SFH, otherwise known as a `parametric SFH'. This assumption has been shown to systematically bias the inferred stellar mass (e.g. \citealt{simha_parametrising_2014,iyer_reconstruction_2017,diemer_log-normal_2017,ciesla_sfr-m_2017}), and has resulted in a more modern approach of modeling SFHs flexibly (i.e. the ``non--parametric method") in a piecewise manner (e.g. \citealt{iyer_reconstruction_2017,leja_how_2019,lower_how_2020,wang_population_2025}).   Similarly, to model dust in SED fitting, it is common to assume a dust attenuation law from a dust grain size distribution and dust geometry; often researchers simply assume a Milky Way dust law, and a uniform `dust screen' model that attenuates all the starlight equally. Unsurprisingly, many combinations of star-dust geometry and varying dust properties can complicate this picture in reality \citep{narayanan_theory_2018,lower_how_2022,hahn_inhomogeneous_2025,mckinney_modeling_2025}. Moreover, stellar models have many systematic uncertainties that will affect the results from SED fitting (e.g. \citealt{robotham_progeny_2024,bellstedt_progeny_2025}).

Outside of many relevant systematic biases that can affect inferred galaxy properties, SED fitting traditionally quotes factors of a few uncertainty in galaxy properties. This said, these stated uncertainties from SED fitting may not always be representative of the true error in the inferred physical properties because of complex degeneracies or systematics in the modeling (e.g. \citealt{acquaviva_simultaneous_2015}). It is therefore critical to our understanding of galaxy formation and evolution to not only understand biases in the inference of galaxy physical properties via SED fitting, but also derive accurate error bars from this analysis.

Attempting to backwards-model galaxy properties is a complex multivariate process that requires careful implementation. Machine learning (ML) techniques are being used across astronomy to study the Universe at many different physical size scales, from exoplanets to cosmology. Work with ML models in the study of galaxies includes but is not limited to image classification, source classification, photometric redshift estimation, and galaxy property estimation (e.g. \citealt{do_galaxiesml_2024,zeraatgari_machine_2024,janiurek_testing_2024,gilda_mirkwood_2021}). Given the powerful potential of ML to approach these inverse problems, it is no surprise that researchers have attempted to apply ML to derive galaxy physical properties. Models of this nature need to be able accept observed light as an input and derive galaxy properties, mirroring the problem/goal structure of traditional SED fitting. However, these approaches directly link observables to physical properties through a model trained by some trusted set of reference data rather than building a physical model each time to match observed spectra. In addition to potentially reducing or removing some of the assumptions that are necessary for traditional SED fitting software, utilizing ML models can also result in a significant speed-up for inference of galaxy properties. 

ML models designed to recover physical properties can be constructed using the results from high-quality real observations as training data or photometry generated from a simple galaxy model using a stellar population synthesis tool such as \textsc{fsps} \citep{conroy_propagation_2009,conroy_propagation_2010} to generate large numbers of synthetic spectra as training data. However, each method comes with its own concerns - observational data represent the real universe, but the training data do not necessarily have a well-known `ground truth' to reference, while constructed synthetic spectra have a fully known ground truth to reference, but they may not represent the complexity of real galaxies or a particular model may not accurately reflect the uncertainty in galaxy formation physics. 

Many groups have developed their own versions of these models that process input photometric or spectroscopic information into inferred galaxy properties. Some of these models are trained on observational data, and others are constructed from simulations for implicit likelihood or simulation-based inference (SBI) strategies. For example, \cite{abedini_cosmos-web_2025} used self-organizing maps to derive the galaxy properties in the COSMOS-WEB survey. For the EUCLID survey, various ML models have been constructed and tested on mock catalogs to eventually run on the real survey, albeit with differing training data and algorithms \citep{euclid_collaboration_euclid_2024,euclid_collaboration_euclid_2025,euclid_collaboration_euclid_2025-1}. \cite{zeraatgari_exploring_2024} constructed a model to infer galaxy stellar mass, star formation rate, and metallicity for low-redshift galaxies trained on photometric data from the SDSS DR8. \cite{zhong_galaxy_2024} develop a model using an autoencoder that can estimate redshifts, classify objects, and reconstruct spectra. \cite{chu_galaxy_2024} use forward-modeled \textsc{IllustrisTNG} images and detailed velocity maps to train a convolutional neural network designed to predict galaxy stellar mass and investigate which inputs are most relevant for successful inference. \cite{ginolfi_inferring_2025} constructed a data pipeline trained on simulated spectra generated from simulated lightcone data to predict the properties of $z\sim1-3$ galaxies. \cite{raghav_photometric_2024} test variety a of ML model types on SDSS DR7 to infer galaxy star formation rates and find promising results. \cite{iglesias-navarro_simulation-based_2025} construct a model that can perform pixel-based SED fitting on simulated JWST spectra at orders of magnitude higher speeds than traditional modeling. \cite{wang_sbi_2023} present an SBI model trained on model galaxies' JWST and Hubble (HST) photometry that can process missing photometric information with a kernel density estimator (KDE) used to draw from the set of instances in the training set that best mirror the input. \textsc{se3d} \citep{zhang_se3d_2025,ramnichal_se3d_2025} instead more closely mirrors traditional SED fitting by modeling galaxy spectra with a trained emulator based on radiative transfer run on simple galaxies. \textsc{synference} \citep{harvey_flexible_2025} is a SBI galaxy inference tool using \textsc{ltu-ili} \citep{ho_ltu-ili_2024} trained to recover galaxy properties with HST and JWST photometry generated with \textsc{synthesizer} forward models of 8 parameter galaxy spectra \citep{lovell_synthesizer_2025,roper_synthesizer_2026} .

While these studies all were highly successful in deriving physical properties of galaxies from particular images or spectra with MK models, they are not designed to be generalized software for SED fitting. Their success suggests that it may be possible to construct a more general model that can derive galaxy properties from arbitrary observed photometry similar to traditional SED fitting. \cite{gilda_mirkwood_2021} made some of the first steps towards a general modern ML-based SED fitting software with their pilot \textsc{mirkwood} model which significantly outperforms traditional SED fitting with the same available input data and also produces output uncertainties that reflect the true uncertainty. Different from most of previously described models, they used a sample of synthetic SEDs generated directly from 3D radiative transfer run on cosmological simulations of galaxies with a variety of galaxy formation models. The training dataset used for \textsc{mirkwood} avoids the pitfalls that can be associated with the models described above; it has reliable ground truth physical properties derived directly from the simulation particles and spectra which come from a 3D realization of a galaxy. However, the pilot model restricted itself to a sample of galaxies from the \textsc{IllustrisTNG}, \textsc{eagle}, and \textsc{simba} simulations specifically at $z=0$. Furthermore, a critical detail lacking from their pilot model, and most of the aforementioned studies, is the ability to process variable amounts of available photometric data, as any individual observation is not guaranteed to have any particular photometric filter available. Most supervised ML techniques are designed to learn a connection between input features and output labels but cannot handle the case where one or many input features are missing.  In other words, they cannot handle the necessary task of evaluating observations that are missing data at even one of the wavelengths the ML model was trained on.  

In this paper, we build on the success of previous work in utilizing machine learning models to determine the physical properties of galaxies and expand it to a general package designed for potential public use that is capable of processing variable input photometric observations and deriving key galaxy physical properties and redshift. We use large suites of cosmological galaxy simulations and 3D radiative transfer run on them to construct our training set. We present our fiducial model \modelnamec, illustrate its success in predicting galaxy properties and appropriate uncertainties, and compare to traditional SED fitting software. We investigate for systematic biases and the ability of \modelname to generalize to observational data by evaluating its performance on novel simulation data.

\section{Methodology}\label{sec:meth}

In this section, we describe the details of our methodology to construct \modelname and compare to standard SED fitting techniques. The goal of \modelname is to be able to infer galaxy properties from photometric observations by referencing the results of cosmological galaxy simulations where we can derive the ground truth physical properties. To accomplish this goal, we forward model simulations run with a variety of galaxy formation physics with radiative transfer. To bridge the inverse problem of connecting observed photometry with redshift and properties, we use machine learning algorithms. Because ML algorithms are often designed to only handle fixed numbers of inputs, we use imputing to address the gap between arbitrary input observations and the trained ML algorithms. 

In Sections~\ref{subsec:train} and \ref{subsec:training}, we describe how we forward model sets of cosmological simulation outputs to a set of reference spectra to train our ML models. In Section~\ref{subsec:model}, we describe in detail the steps by which \modelname converts input photometry to inferred galaxy properties with uncertainties. Finally, in Section~\ref{subsec:prosp} we describe the setup by which we compare the performance of \modelname and the industry standard observational software packages used to perform SED fitting.

\subsection{Training Data}\label{subsec:train}
In this subsection, we outline the strategy we used to generate our model training data. Our goals with the data we use to construct our SED modeling software are to (1) have self-consistently generated model galaxies, (2) not be tied to the assumptions of any particular galaxy formation physics model, (3) represent the uncertainty in these formation models in our training set, and (4) represent the galaxy simulations with realistic mock photometry. To fulfill (1), we elect to use galaxies generated from cosmological hydrodynamic galaxy simulations. To satisfy (2) and (3), we utilize the results from the 1-parameter (1P) suite of the Cosmology and Astrophysics with Machine Learning Simulations (CAMELS) suite for \textsc{simba} and \textsc{IllustrisTNG} \citep{villaescusa-navarro_camels_2021,villaescusa-navarro_camels_2023,ni_camels_2023}. Finally, to ensure realistic photometry, we run 3D radiative transfer on the simulated galaxies.

\subsubsection{Cosmological Simulations}\label{subsec:sims}

To marginalize over the uncertainty in our galaxy formation physics models, we aggregate results from the expanded CAMELS 1P simulation suite for \textsc{IllustrisTNG} and \textsc{simba}. The CAMELS 1P suite consists of numerous simulation volumes of these galaxy formation models run to $z=0$ with the physics model parameters varied. In particular, the expanded CAMELS 1P \citep{ni_camels_2023} set contains boxes from varying 28 cosmological and astrophysical parameters in each simulation one at a time. We focus only on CAMELS 1P physics variations where astrophysical parameters are varied; therefore, we do not assume that any particular physical model is the `correct' physical representation of the Universe nor do we tie ourselves to the fiducial physics parameters of the simulations. Almost every astrophysical parameter is varied in total four times from the fiducial value - twice below the fiducial value and twice above. Below, we briefly describe the \textsc{simba} and \textsc{IllustrisTNG} physics models.

The \textsc{simba} galaxy formation model \citep{dave_simba_2019} runs on the \textsc{gizmo} hydrodynamic code \citep{hopkins_new_2015} and is the successor to the \textsc{mufasa} simulations \citep{dave_mufasa_2016}. \textsc{gizmo} is based on the \textsc{Gadget-3} hydrodynamic code \citep{springel_cosmological_2005}. The \textsc{simba} galaxy formation model star formation rate (SFR) is based on the local $\text{H}_2$ density calculated based on the \citet{krumholz_comparison_2011} model, the dynamical time, and an assumed efficiency of 2\% \citep{kennicutt_global_1998}. Star particles follow a Chabrier initial mass function (IMF) \citep{chabrier_galactic_2003} and enrich the ISM based on the yield models from \cite{nomoto_nucleosynthesis_2006}, \cite{iwamoto_nucleosynthesis_1999}, and \cite{oppenheimer_cosmological_2006} for Type II SNe, Type 1a SNe, and AGB stars respectively. \textsc{simba} tracks 11 elements over cosmic time. The gas cooling is calculated by the \textsc{grackle} chemistry module \citep{smith_grackle_2017} with the self-shielding model of \cite{rahmati_evolution_2013} against an ionizing background calculated from the \cite{haardt_radiative_2012} model. Stellar feedback is present in metal-loaded winds with mass-loading factors based on the results of \cite{angles-alcazar_cosmic_2017}. \textsc{simba} additionally includes a model for supermassive black holes (SMBHs). They are seeded at $M_{\rm BH}=10^4M_{\odot}/h$ once a galaxy reaches a stellar mass threshold and grow based on two modes: a cold accretion mode based on \cite{angles-alcazar_gravitational_2017} and a hot halo mode based on \cite{bondi_spherically_1952}. Black hole feedback similarly comes in both kinetic and radiative modes; kinetic feedback is jet-based for low accretion and wind-based for high accretion, while radiative feedback is based off of an X-ray model dumping energy into gas \citep{choi_radiative_2012}. 

Relatively uniquely, \textsc{simba} features a self-consistent dust production, growth, and destruction model \citep{dave_simba_2019,li_dust--gas_2019}. All \textsc{simba} dust grains are assumed to have the same constant size of 0.1 $\mu$m and are tied to the gas particles. Initial dust production is modeled by by assuming fixed fractions of the stellar metal yields are in dust \citep{dwek_evolution_1998}. The condensation involves the newer models of \cite{ferrarotti_composition_2006} and \cite{bianchi_dust_2007}.
The dust grains are also allowed to `grow' by accreting local gaseous metals. Dust can be destroyed by thermal sputtering, a subgrid prescription for supernova shocks \citep{mckinnon_dust_2016}, AGN jet feedback \citep{angles-alcazar_gravitational_2017}, AGN X-ray emission \citep{choi_radiative_2012}, winds, and star formation. This model is tuned to reproduce the $z=0$ dust mass function and is successful in reproducing higher redshift dust measurements \citep{li_dust--gas_2019}.  For the \textsc{simba} models included in this paper, the dust content seen in the radiative transfer is produced self-consistently in the hydrodynamic simulation, assuming a \citet{weingartner_dust_2001} grain size distribution.

\textsc{IllustrisTNG} \citep{weinberger_simulating_2017,pillepich_simulating_2018,pillepich_first_2018,naiman_first_2018,springel_first_2018,nelson_first_2018,marinacci_first_2018} runs on the \textsc{Arepo} hydrodynamic code \citep{springel_moving-mesh_2011} and is the successor to the \textsc{Illustris} model \citep{vogelsberger_properties_2014,vogelsberger_introducing_2014,genel_introducing_2014,sijacki_illustris_2015}. Star formation in the \textsc{IllustrisTNG} is stochastic for gas particles above the number density threshold $n > 0.13$ cm$^{-3}$, and gas particles are also artificially pressurized by a prescription for an effective equation of state \citep{springel_cosmological_2003}. The \textsc{IllustrisTNG} model assumes a Chabrier IMF \citep{chabrier_galactic_2003}. Star particles enrich the ISM based on yield tables from \cite{portinari_galactic_1998} and \cite{kobayashi_galactic_2006} for Type II SNe, \cite{nomoto_nucleosynthesis_1997} for Type Ia SNe, and \cite{karakas_updated_2010}, \cite{doherty_super_2014}, and \cite{fishlock_evolution_2014} for AGB stars. \textsc{IllustrisTNG} tracks 9 elements explicitly and a tenth `other metals' category. The gas cooling follows a combination of lookup tables and an on-the-fly model from \cite{katz_cosmological_1996} and \cite{wiersma_effect_2009}, an evolving uniform ionizing background as described in \cite{faucher-giguere_new_2009}, and the same hydrogen self-shielding model from \cite{rahmati_evolution_2013} as in \textsc{simba}. Stellar feedback winds are ejected isotropically from their source gas cell as a mix of kinetic and warm thermal feedback with redshift-dependent wind velocities and metallicity-dependent energies. \textsc{IllustrisTNG} also includes a novel magnetic field model.

\textsc{IllustrisTNG} SMBHs are seeded at $8\times10^5 M_{\odot}/h$ for DM halos that meet the threshold $M>5\times10^{10} M_{\odot}/h$. SMBHs in the \textsc{IllustrisTNG} model can grow through \cite{bondi_spherically_1952} accretion up to the Eddington limit. The SMBH feedback falls into several modes as a function of the current fraction of the Eddington limit that depends on the current BH mass \citep{weinberger_simulating_2017}. In the low-accretion state, the model injects kinetic energy as feedback in momentum-based winds, while in the high-accretion state, the energy is injected into the surrounding medium as radiative feedback.

All CAMELS simulated cosmological volumes have $25/h$ Mpc sides in comoving coordinates and contain $256^3$ particles and a mass resolution of $2.3\times10^6 M_{\odot}$ and $1.2\times10^7 M_{\odot}$ for gas and dark matter particles respectively. As a trade-off to explore the parameter space of simulation physics, this resolution corresponds to the resolution typically used for larger, lower resolution runs of these simulations. The default cosmology used in CAMELS volumes is flat, with $H_0 = 67.11 \text{ km/s/Mpc}$, $\Omega_b=0.049$, $n_s=0.8625$ and $\Omega_{m,0}=0.3$.

\subsubsection{Simulated Galaxies Sample Selection}\label{subsubsec:simgals}

We identify galaxies within the CAMELS 1P \textsc{simba} simulations at the snapshots corresponding to integer redshifts $z = 0-6$ using \textsc{caesar} \citep{thompson_pygadgetreader_2014}, a package utilizing \textsc{yt} \citep{turk_yt_2011} and a 6D friends-of-friends (FOF) algorithm to group galaxies and dark matter halos. A `galaxy' is defined as a group of bound particles in the simulations featuring a minimum of 24 star particles. \textsc{caesar} also computes relevant quantities such as the stellar mass and dust mass of the identified galaxies.

For \textsc{IllustrisTNG}, galaxies are identified with the \textsc{subfind} FOF algorithm \citep{springel_simulations_2005}. Unlike \textsc{caesar}, \textsc{subfind} groups particles into the local dark matter subhalos that they are a part of rather than based on the groups of bound star particles. For consistency with the \textsc{caesar} definition and for our resolution cut, we treat any subhalo with 24 or more star particles as a `galaxy'. As for \textsc{simba}, we note and save the information for any galaxies at integer redshifts $z = 0-6$.

\subsubsection{Radiative Transfer and SED Processing}\label{subsubsec:simrad}

\begin{table*}
    \centering
    \begin{tabular}{|c|c|c|c|c|}
        \hline
        {\bf Model} & {\bf 1P Parameter \#} & {\bf Brief Description} & {\bf Significantly affects $z=0$ SMHM?} \\
        \hline
        \textsc{simba} & 3 &  SN mass & N\\\hline

        \textsc{simba} & 4  & AGN mass & N\\\hline

        \textsc{simba} & 5  & SN wind speed & N\\\hline

        \textsc{simba} & 6 & AGN jet speed & N\\\hline

        \textsc{IllustrisTNG} & 3 & winds energy & N\\\hline
        
        \textsc{IllustrisTNG} & 4 & AGN kinetic mode energy & N\\\hline

        \textsc{IllustrisTNG} & 5 & SN wind speed & N\\\hline

        \textsc{IllustrisTNG} & 6 & AGN kinetic mode burstiness & N\\\hline

        \textsc{simba} & 14 & wind recoupling density & N\\\hline

        \textsc{IllustrisTNG} & 17 & wind recoupling density & N\\\hline

        \textsc{simba} & 15 & SMBH seed mass & Y\\\hline

        \textsc{IllustrisTNG} & 22 & SMBH seed mass & N\\\hline

        \textsc{simba} & 16 &  SMBH accretion rate multiplier & Y\\\hline

        \textsc{IllustrisTNG} & 23  & SMBH accretion rate multiplier & N\\\hline

        \textsc{simba} & 17 & SMBH radiative efficiency & Y\\\hline

        \textsc{IllustrisTNG} & 26  & SMBH radiative efficiency & Y\\\hline

        \textsc{simba} & 18  & SMBH accretion rate limiter & N\\\hline

        \textsc{IllustrisTNG} & 24 & SMBH accretion rate limiter & N\\\hline

        \textsc{simba} & 28 & SMBH jet mode threshold & Y\\\hline

        \textsc{IllustrisTNG} & 27 & SMBH quasar mode threshold & N\\\hline

        \textsc{simba} & 11 & SFR efficiency & Y\\\hline

        \textsc{simba} & 22  & wind power-law scaling & Y\\\hline

        \textsc{simba} & 25 & $M_*$ SMBH seeding threshold & Y\\\hline

        \textsc{IllustrisTNG} & 10  & maximum gas consumption timescale & Y\\\hline

        \textsc{IllustrisTNG} & 13 & minimum SN Type II mass & Y\\\hline

        \textsc{IllustrisTNG} & 15 & wind momentum per star formation & Y\\\hline

        \textsc{IllustrisTNG} & 18  & dependence of energy reduction on Z & Y\\\hline

        \textsc{IllustrisTNG} & 19 & Z at which wind energy pivots & Y\\\hline

    \end{tabular}

\caption{\textbf{Table of CAMELS 1P cosmology simulation outputs used to train our models.} \rm{The first column represents which galaxy formation model is being referenced, the second column provides the CAMELS parameter reference number, the third column provides a brief description of the parameter, and the final column notes whether the changes in the parameter were listed by Oh et al. (in prep) as causing the most significant shifts in the $z=0$ stellar mass - halo mass relation amongst the 1P runs.}}
\label{table:1p}
\end{table*}

We employ \textsc{powderday} \citep{narayanan_powderday_2021}, an open-source 3D radiative transfer package that wraps several software packages to form a complete pipeline to convert the raw simulation galaxy outputs identified with \textsc{caesar} and \textsc{subfind} to synthetic spectra. The package \textsc{yt} \citep{turk_yt_2011} directly interfaces with simulation particles to extract their properties and smooth the gas (and dust) onto a grid in the form of an octree and Voronoi mesh for \textsc{simba} and \textsc{IllustrisTNG} respectively. Stellar spectra are generated with \textsc{fsps} \citep{conroy_propagation_2009,conroy_propagation_2010} by assuming that each star particle can be treated as a simple stellar population with age and metallicity defined by the formation time and metallicity saved in the simulation outputs. In \textsc{fsps}, we use the default \textsc{mist} isochrones \citep{paxton_modules_2011,paxton_modules_2013,paxton_modules_2015,dotter_mesa_2016,choi_mesa_2016} and \textsc{miles} spectral library \citep{vazdekis_evolutionary_2010,falcon-barroso_updated_2011,vazdekis_evolutionary_2015} with an assumed Chabrier IMF \citep{chabrier_galactic_2003} to match both of the galaxy formation models in this paper. \textsc{hyperion} \citep{robitaille_hyperion_2011} performs the 3D radiative transfer on the grid by generating photon packets in iterations that travel through the model dust grid and interact randomly. This process is repeated until 99\% of the dust cells have their temperatures change by less than 1\% between successive iterations. Dust extinction laws are held fixed in each cell to be a \cite{weingartner_dust_2001} law, and the normalization is determined by the dust mass in the cell (because of star-dust geometry, the dust \textit{attenuation} law will still vary galaxy to galaxy despite the extinction law being fixed). For \textsc{simba}, dust masses come directly from the current dust mass generated by the dust model. However, \textsc{IllustrisTNG} does not have a dust model. We therefore assume a dust-to-metals (DTM) ratio of 0.4 for \textsc{IllustrisTNG} galaxies. We note that \textsc{IllustrisTNG} galaxies typically have significantly more gas particles in their FOF groups, and so commonly have higher dust masses than those associated with \textsc{simba} galaxies. We do not include black holes as a source of radiation, though they naturally drive the evolution of galaxies in our sample. Polycylcic aromatic hydrocarbons (PAHs) are assumed to be a fixed fraction (5.86\%) of the dust mass.

We run \textsc{powderday} on every galaxy for the CAMELS 1P suite for \textsc{simba} and \textsc{IllustrisTNG} at integer redshifts from $z=0-6$ for a specific set of 1P varied astrophysical parameters in addition to the cosmological boxes that use the fiducial \textsc{simba} and \textsc{IllustrisTNG} physics. In total, this creates a sample of $\sim$300,000 synthetic SEDs from simulated galaxies. Specifically, amongst the 28 astrophysical and cosmological parameters that the CAMELS 1P suite varies in the \textsc{simba} and \textsc{IllustrisTNG} models, we only select galaxies from cosmological volumes whose varied astrophysical parameter either has an analog in both galaxy formation models (e.g. black hole seed mass) or 
are noted by Oh et al. (in prep) to have significant impact on the z=0 stellar-mass-halo-mass relation. We caution the reader that even if a parameter does have an analogous term in the other simulation with an identical physical interpretation, it does not necessarily mean that it is implemented the same way or that varying it will have the same impact on the evolution of the simulation. In Table~\ref{table:1p}, we explicitly list these 1P astrophysical parameters corresponding to the portion of the 1P sample that we use in training of our model and describe their physical meaning.

\begin{table*}[ht]
    \centering
    
    \begin{tabular}{|c|c|c|c|c|c|}
        \hline
        {\bf Instrument} & {\bf Filter} & {\bf $\lambda_{\rm eff}$ ($\mu$m}) & {\bf Instrument} & {\bf Filter} & {\bf $\lambda_{\rm eff}$ ($\mu$m})\\
        \hline
 
        GALEX & FUV & 0.1528 & JWST MIRI &  F560W & 5.6115\\\hline
        GALEX & NUV & 0.2271 & JWST MIRI &  F770W & 7.591 \\\hline
        HST WFC3 & F275W & 0.2700 & JWST MIRI &  F1000W & 9.923\\\hline
        HST WFC3 & F336W & 0.3347 & JWST MIRI &  F1280W & 12.77\\\hline
        HST WFC3 & F475W & 0.4736 & JWST MIRI &  F1500W & 15.01\\\hline
        HST WFC3 & F555W & 0.5256 & JWST MIRI &  F1800W & 17.94\\\hline
        HST WFC3 & F606W & 0.5813 & JWST MIRI &  F2100W & 20.70 \\\hline
        HST WFC3 & F814W & 0.7973 & WISE & W1 & 3.346\\\hline
        HST WFC3 & F105W & 1.048 & WISE & W2 & 4.595\\\hline
        HST WFC3 & F110W & 1.136 & WISE & W3 & 11.55\\\hline
        HST WFC3 & F125W & 1.241 & WISE & W4 & 22.08\\\hline
        HST WFC3 & F140W & 1.383 & Spitzer MIPS & 24 $\mu$m & 23.43\\\hline
        HST WFC3 & F160W & 1.532 & Herschel PACS & 70 $\mu$m & 70.02 \\\hline
        SDSS & u & 0.3546 & Herschel PACS & 100 $\mu$m & 99.62\\\hline
        SDSS & g & 0.4670 & Herschel PACS & 160 $\mu$m & 158.6\\\hline
        SDSS & r & 0.6156  & Herschel SPIRE & 250 $\mu$m & 246.1 \\\hline
        SDSS & i & 0.7472 & Herschel SPIRE & 350 $\mu$m & 345.5\\\hline
        SDSS & z & 0.8917 & Herschel SPIRE & 500 $\mu$m & 493.1\\\hline
        JWST NIRCam &  F070W & 0.7014 \\\hline
        JWST NIRCam &  F090W & 0.8983 \\\hline
        JWST NIRCam &  F115W & 1.1486 \\\hline
        JWST NIRCam &  F150W & 1.4944 \\\hline
        JWST NIRCam &  F200W & 1.9782 \\\hline
        JWST NIRCam &  F277W & 2.7605 \\\hline
        JWST NIRCam &  F356W & 3.5484 \\\hline
        JWST NIRCam &  F444W & 4.3787 \\\hline
 
    \end{tabular}
\caption{\textbf{List of photometric filters we derive from our synthetic \textsc{Powderday} spectra.} \rm{Columns 1 and 4 list the instrument, columns 2 and 5 list specific photometric filters, and columns 3 and 6 state the effective wavelength for the photometric filter. We compute photometry in the UV to the far-IR. We access the filter transmission curves and integrate with \textsc{sedpy} \citep{johnson_bd-jsedpy_2021}.}}
\label{table:filters}
\end{table*}

\begin{figure*}[ht]
    \epsscale{1.2}
    \plotone{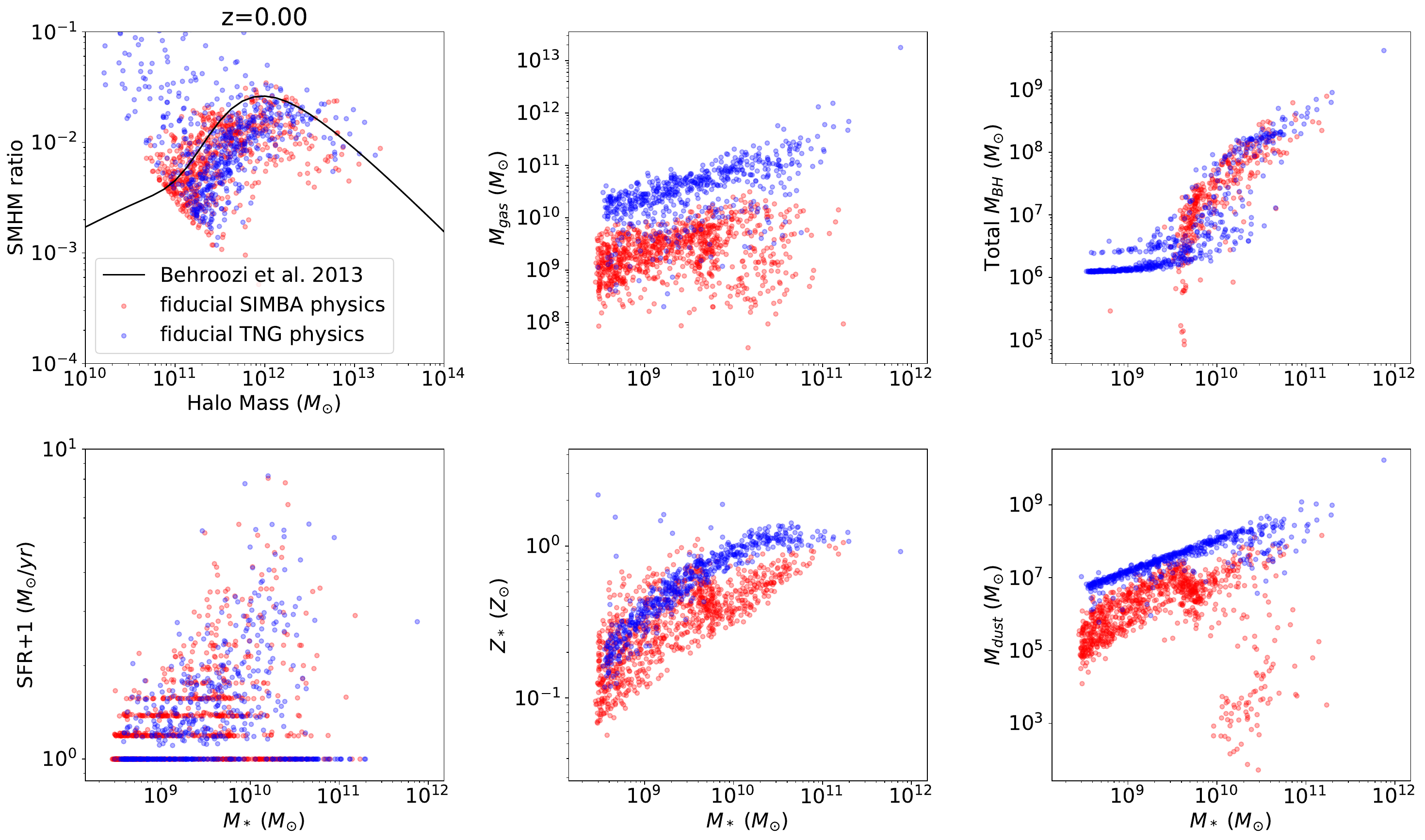}
    \caption{\textbf{Summary of fiducial physics \textsc{Simba} and IllustrisTNG galaxies' physical relations at $z=0$.} Top left: The stellar mass - halo mass relation for \textsc{simba} and \textsc{IllustrisTNG} galaxies compared to \cite{behroozi_lack_2013}. Top middle: The stellar mass - gas mass relation for galaxies. Top right: The stellar mass - black hole mass relation for galaxies. Bottom left: the stellar mass - star formation rate relation for galaxies. Bottom middle: The stellar mass - metallicity relation for galaxies. Bottom right: Stellar mass - dust mass relation, where the dust mass for \textsc{simba} galaxies is derived from the simulation's dust model and the \textsc{IllustrisTNG} dust mass is derived from assuming a DTM ratio of 0.4.}
    \label{fig:sim_sum_grid}
\end{figure*}

\begin{figure*}[hb]
    \plotone{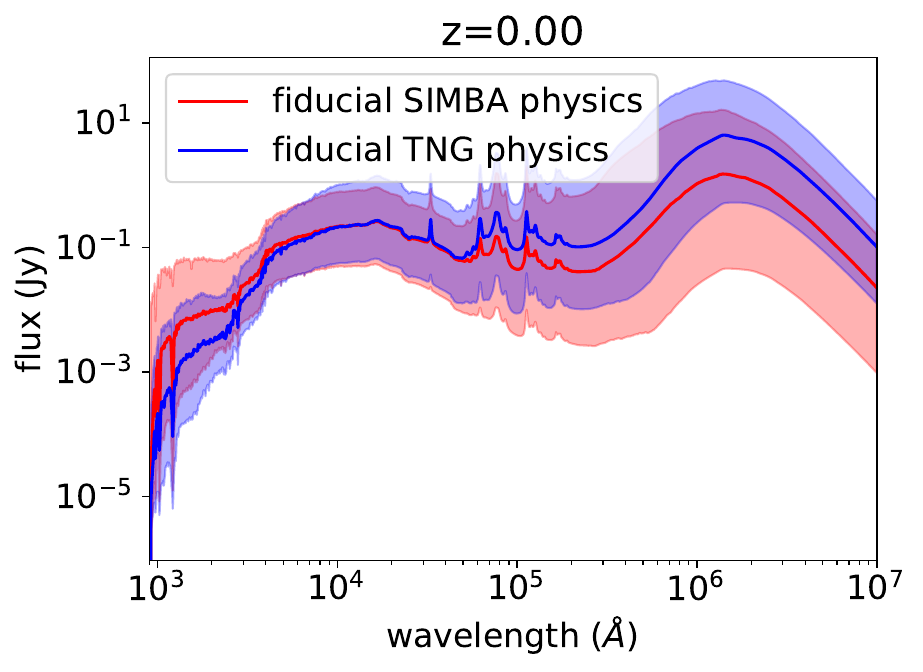}
    \caption{\textbf{Distribution of spectra for \textsc{Simba} and IllustrisTNG galaxies generated by 3D radiative transfer at $z=0$.} Solid lines represent the median emission at a particular wavelength amongst all the $z=0$ spectra and the shaded regions represent the 16th and 84th percentiles of the distribution of the spectra. }
    \label{fig:sim_sum_sed}
\end{figure*}

\begin{figure*}[ht]
    \plotone{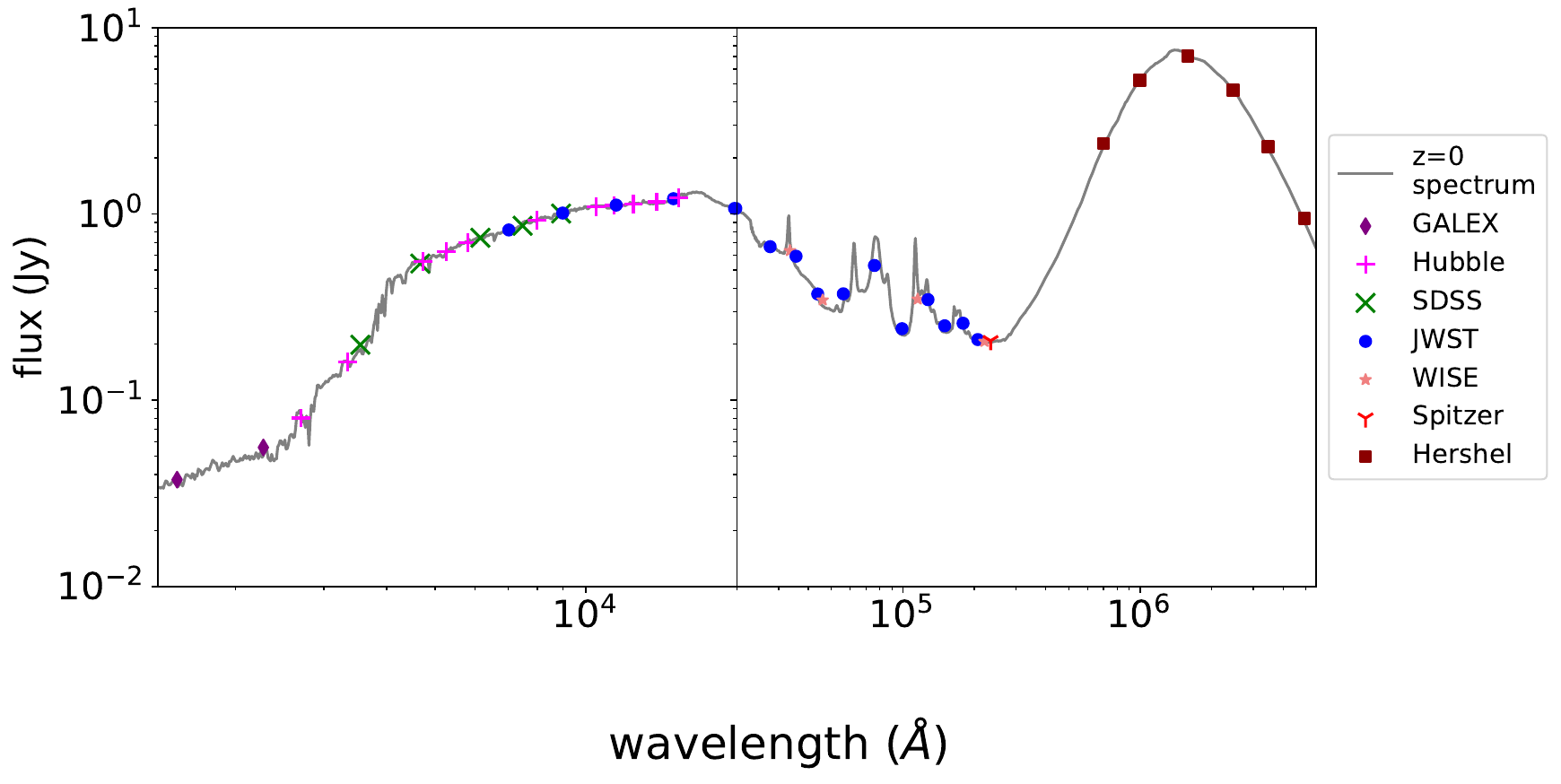}
    \caption{\textbf{Visual representation of the photometric filters we train \modelname with} for an example $z=0$ spectrum. The left half plot is at a different scale than the right half to better present the filters visually up to $\sim2$ $\mu$m.}
    \label{fig:sed_filt}
\end{figure*}

This broad sample enables us to construct a catalog of mock galaxy SEDs across different epochs in these simulations. We compute photometry for these SEDs by integrating the redshifted spectra in a set of photometric filters.  We list the filters we use in this work in Table~\ref{table:filters}. For any individual calculation, we redshift a spectrum by an integer multiple of 0.1 plus a random offset $\leq0.05$. This process is repeated for every mock SED from from our sample for the redshift their galaxy snapshot was at up to $z_{\rm snap}\pm1$. We use \textsc{sedpy} \citep{johnson_bd-jsedpy_2021} to integrate the redshifted spectra in the appropriate photometric bands. The combination of the photometry calculated at many different redshifts and the random offset allows the training set to contain a continuous redshift distribution rather than having the luminosity distances only be at fixed, discrete values.
Every galaxy that we are able to successfully run \textsc{powderday} on is included in the training or testing set. We compute galaxy properties from a combination of the catalog outputs and the \textsc{powderday} grid properties. To illustrate what the sample of data looks like, we plot summary relations for the CAMELS boxes with the fiducial \textsc{simba} and \textsc{IllustrisTNG} physics at $z=0$ as well as the distributions of spectra in Figures~\ref{fig:sim_sum_grid} and \ref{fig:sim_sum_sed}. We briefly note a few interesting features in Figure~\ref{fig:sim_sum_grid} - at fixed stellar mass, a typical \textsc{IllustrisTNG} galaxy has more gas than a \textsc{simba} galaxy. \textsc{simba} has a larger spread of metallicity values at lower stellar masses, and lower metallicities at higher stellar masses. In the bottom right panel, the \textsc{simba} dust masses come directly from the dust model, whereas the \textsc{IllustrisTNG} value comes from the assumed dust-to-metals ratio. Given that we assign dust to \textsc{IllustrisTNG} based off a DTM ratio in gas, it is not surprising that with more gas and a tighter relation between stellar mass and metallicity, that there is a tight stellar mass - dust mass relation with higher normalization than \textsc{simba} at fixed stellar mass. In Figure~\ref{fig:sim_sum_sed}, we illustrate the distribution of spectra for the fiducial galaxy physics boxes; \textsc{IllustrisTNG} galaxies are typically more attenuated, likely because of their greater dust masses at fixed stellar mass, whereas there is more spread in \textsc{simba} in the dust emission, likely because there is a broader distributiion of dust masses at fixed stellar mass. In Figure~\ref{fig:sed_filt}, we also visualize where on the SED each of our computed filters lie for a $z=0$ SED. We include two axis scales in the figure to make the filters between $0.15-1 $ $\mu$m easier to recognize.

\subsection{Training Set}\label{subsec:training}
From the CAMELS galaxy catalogs and radiative transfer, we have a large sample of galaxies generated with a variety of implemented galaxy physics. The next concern is how to divide them into a training set for our ML models. In principle, we could construct our model to predict an arbitrary number of physical properties from photometry. In this paper, we present our fiducial \modelname version trained to infer galaxy stellar mass ($\log{(M_*/M_{\odot})}$), dust mass ($\log{(M_{\rm{d}}/M_{\odot})}$), and average star formation rate in the last 100 Myr ($\log{((\rm{SFR}_{100}+1)/(M_{\odot}/\rm{yr})})$) -- some of the most standard physical properties that are commonly quoted from SED fitting results. In constructing the fiducial model, we allow only photometry corresponding to a galaxy's intrinsic redshift $\pm1$ to be in the training set. This choice is intended to limit the effect that model galaxies can have at redshifts significantly different from the simulation snapshot they correspond to. However, the training set still retains the model galaxies resulting from the different evolutionary histories corresponding to the varied astrophysical parameters in the CAMELS 1P suite. 
To train our fiducial models, we randomly assigned 70\% of our galaxies to the training set and 30\% to the testing set. The photometry corresponding to any individual galaxy is confined entirely to the either the training or testing set; this step guarantees that at the testing phase any photometry being evaluated represents a `novel' galaxy to \modelnamec.

\subsection{Model}\label{subsec:model}

In this subsection, we describe the components of \modelname that allows it to make inferences about galaxy properties. There are 3 main components to producing predictions: (1) Monte Carlo (MC) photometric sampling, which samples the uncertainties of the input photometry, (2) imputation, which fills in missing photometry with close analogs in the training set to standardize input size for the remaining steps, and (3) photometric-z (hereafter photo-z) estimation and physical property inference tied to ML models. We describe each of these in turn below.

{\it Photometry Uncertainties:} \modelname first receives a set of observed photometric values and their uncertainties as an input for a galaxy observation. We treat each input photometric value and its uncertainty as the main and standard deviation of a Gaussian distribution and draw from the distribution N times. The uncertainty is based on an input signal-to-noise ratio (SNR). This generates a set of potential photometric shapes that reflect the uncertainty in each filter. 

{\it Imputation:} In order to deal with ``missing data'', i.e. when the observer does not have access to all the exact photometric bands that we train on, we employ a technique known as imputation.  
Imputation is a statistical technique whereby rather than ignoring data that do not have a complete set of features, one attempts to infer what the missing values might be.
More sophisticated imputation techniques essentially turn the missing data problem into a regression problem.
For \modelnamec, we elect to use a K-nearest neighbors (KNN) imputation strategy. This strategy involves filling in filters for an input observation with a reference set of complete information; it implicitly assumes that if an instance is similar in some number of features to a reference instance, it is likely similar in the remaining features. For any input with missing data, the code will compute some notion of the `distance' in feature space to each reference instance. The algorithm will note the K instances with the least evaluated distance and combine their features to produce the imputed value of the missing features for the input instance.  

In \modelnamec, we use the {\sc sklearn} implementation of a KNN imputer \citep{pedregosa_scikit-learn_2011}. Our library of synthetic galaxy photometry produced by radiative transfer (see \ref{subsubsec:simrad}) is used to train the imputer. We have scaled the photometric flux values in the training set by first taking the logarithm of their values and then scaling them to a distribution with a mean of 0 and standard deviation of 1. The distance metric is the Euclidean distance in this scaled feature space. We take the K=5 nearest neighbors determined and average their features weighted by the inverse of their distance to the input instance in the known filters. These choice prevents any particular photometric band on a part of the spectrum from being preferred over another because of typical values in that filter and weights the reference spectrum that best matches the input overall higher. Additionally, we use a KNN imputer because it will produce physically-motivated spectra that resemble the self-consistently produced synthetic spectra it is trained on. Empirically, when we initially tested the performance of KNN imputation relative to other imputation strategies, we found that it both significantly outperformed other imputation strategies and also was the only strategy that returned distributions resembling spectra. We also fix the number of galaxies that are allowed to be in the imputation set at each redshift to avoid an imbalance of galaxy number as a function of redshift affecting the imputer.

Although the KNN imputer does not produce uncertainty in any individual outputs, we can still estimate the uncertainty in the spectrum by comparing the prediction for each of our MC-generated photometric values based off the input photometric values and their uncertainties from the first step of the \modelname workflow. This step will output varied imputed sets of filter values corresponding to the uncertainty in the observations. The imputation importantly method guarantees that for any general observation with an arbitrary number of observed filters, the machine learning step following the imputation will always receive the same number of features.

{\it Photo-z and Physical Property Estimation:} The set of $N$ iterations of MC variation from the photometric observations after imputation are now used to predict galaxy properties. The first necessary component is to evaluate the redshift. We train our photo-z model with our training set of galaxy redshift - photometry pairs to construct an ML model to with the input being the galaxy photometry scaled as in the imputing step to output photo-z predictions. For the particular ML model strategy, we use the {\sc NGBoost} algorithm \citep{duan_ngboost_2020}, a variant of gradient boosting strategies that also outputs probability distributions (a predicted $\mu$ and $\sigma$ of a Gaussian distribution). For each observation, we randomly draw from the distribution to sample the output uncertainties.
The output of the redshift prediction is then subsequently scaled and included as in input for the physical property inference. For each physical galaxy physical property of interest ($M_*,M_{\rm{d}},\rm{SFR}_{100}$), we have trained an additional {\sc NGBoost} model for physical property predictions based on the input photometry and redshift. For each of the $N$ iterations, we also draw a sample from the output {\sc NGBoost} posterior from these models. Finally, the set of predictions for properties is treated as the output posterior for \modelnamec. The median of each distribution and the 16th and 84th percentiles are then treated as the prediction and confidence interval respectively.

\subsection{Comparisons with observational tools}\label{subsec:prosp}

\begin{table*}[ht]
    \centering
    \begin{tabular}{|c|c|c|c|c|}
        \hline
        {\bf Parameter} & {\bf Fixed?} & {\bf Value/Prior} & {\bf Description}\\
        \hline
 
        IMF & Y & \cite{chabrier_galactic_2003} & galaxy IMF\\\hline

        Dust type & Y & \cite{calzetti_dust_2000} & diffuse dust attenuation applied to all stars\\\hline

        dust2 & Y & Uniform $[0.0,10.0]$ & diffuse dust attenuation value \\\hline

        add\_agb\_dust\_model & Y & False & include AGB star dust model \\\hline
        
        duste\_gamma & N & Uniform $[0.0,1.0]$ & fraction of dust heated at minimum radiation field\\\hline

        duste\_umin & N & Uniform $[0.1,25.0]$ & minimum radiation field in MW units\\\hline

        duste\_qpah & N & Uniform $[0.0,10.0]$ & percentage of dust mass in PAHs\\\hline

        logmass & N & Uniform $[7,14]$ & log galaxy formed stellar mass in $M_{\odot}$\\\hline

        logzsol & N & Uniform $[-3,1]$ & log galaxy metallicity in $Z_{\odot}$\\\hline

        SFH history & Y & Dirichlet \citep{betancourt_hamiltonian_2013} &  $M_*$ fraction per time bin, first bin is last 100 Myr\\\hline 

        z & Y/N & Uniform $[0,8]$ & redshift\\\hline
 
    \end{tabular}

\caption{\textbf{Table of \textsc{Prospector} priors used when fitting synthetic galaxy photometry.} \rm{We assume a Dirichlet concentration parameter $\alpha=1.0$. For the fits where \textsc{prospector} has all available photometric data, redshift is fixed to its true value, and in our test cases where there is missing photometry, it is allowed to vary.}}
\label{table:prosp_prior}
\end{table*}

Throughout our results section, we compare the performance of \modelname in recovering important galaxy physical properties to the traditional SED fitting tool \textsc{prospector} \citep{leja_deriving_2017,leja_how_2019,johnson_stellar_2021}. \textsc{prospector}  uses \textsc{fsps} \citep{conroy_propagation_2009,conroy_propagation_2010} for stellar population synthesis, which is the exact same model that \textsc{powderday} uses for stellar spectra generation, and uses the \textsc{dynesty} code \citep{skilling_nested_2004,skilling_nested_2006,higson_dynamic_2019,speagle_dynesty_2020,sergey_koposov_joshspeagledynesty_2025} to explore the model parameter space.  We set the stellar spectral library and isochrone choices the same in both {\sc prospector} and {\sc powderday} to avoid any systematic uncertainty that would otherwise be introduced by a mismatch in spectral libraries or isochrones, though note that this systematic would exist for real observations.

Given the wide variety of galaxy properties, growth histories, and redshifts that are being fit in the test set, we run \textsc{prospector} with broad priors such that it has flexibility to fit the photometry on hand. This decision is especially relevant for parameters related to dust attenuation and emission since our SEDs include star-dust geometry effects from radiative transfer. We model galaxies with a non-parametric Dirichlet SFH \citep{betancourt_hamiltonian_2013}, where most recent bin is fixed to be the last 100 Myr. The remainder of the bins are logarithmically spaced in time. In Table~\ref{table:prosp_prior} below, we explicitly list the other priors and parameters we use when running \textsc{prospector} in this work.

We run \textsc{prospector} only on galaxy photometry corresponding to the test galaxies' true intrinsic redshift (e.g. a galaxy from a $z=1$ snapshot is only fit by \textsc{prospector} for its $z=1$ photometry) to allow for self-consistent comparisons with the star formation histories. We do not evaluate galaxies at $z=0$ because the random offset applied to redshift when integrating the SED to photometric bands combined with the physical lower bound of $z=0$ causes galaxies at $z=0$ to be sparsely represented in the training set, and therefore it is challenging for \modelname to fit these spectra. For the ideal test case where all the photometry is available to constrain the galaxy properties, we fix the redshift in \textsc{prospector} fits to its true value, whereas for the fits with limited photometric constraints, we treat redshift as a free parameter in \textsc{prospector}. Though it is not standard practice to allow the \textsc{prospector} redshift to be treated as free when fitting, we want the uncertainties on physical properties that \textsc{prospector} outputs to be comparable to \modelnamec, where the redshift uncertainty is propagated through the process. We also separately run the standard tool \textsc{eazy-py} \citep{brammer_eazy_2008} in these cases to reflect typical strategies of inferring the photometric redshift through photo-z codes separately before the SED fitting process and to compare performance to our photo-z step.  We note that we assume a signal-to-noise ratio (SNR) of 10 on all photometric filters when running all SED modeling packages throughout the text.

\section{KNN Imputer Evaluation}\label{sec:KNN}

\begin{figure*}[htp]
    \epsscale{0.8}
    \plotone{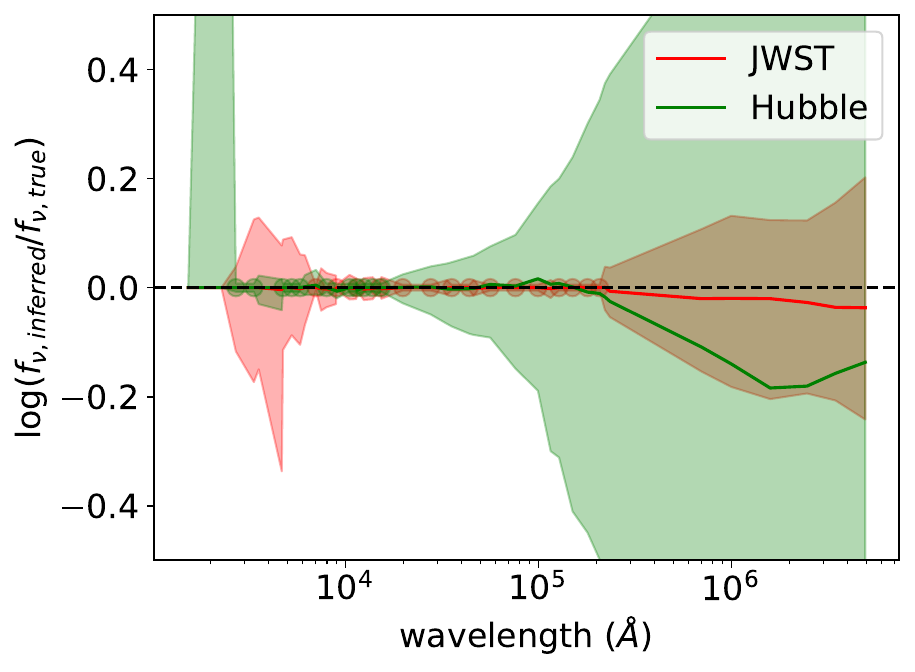}
    \caption{\textbf{Demonstration of the performance of the KNN imputer that we include in \modelname to handle missing data.} This test evaluates whether the KNN imputer is successfully reconstructing the shape of the galaxy SEDs based on the available photometry. The KNN imputer is trained with access to the complete set of photometry, and then we take the ratio of the SED the imputer typically recovers with access to only the JWST or HST bands to the true SED. A bias in the inferred SEDs would affect the downstream inference of galaxy properties based on these inferred SEDs. The colored points reflect the values of photometry that are known for the KNN imputer (red for JWST, green for HST). The solid lines reflect the median performance and the shaded regions correspond to the 16th and 84th percentiles of the ratios between the imputed flux and the known true flux for the test sample. The solid median lines remaining near the 1-to-1 line across the range of the spectrum suggests the KNN imputer is successfully recovering the initial spectrum.}
    \label{fig:knn_imp}
\end{figure*}

Before making detailed comparisons of \modelname's performance to the predictions of traditional SED fitting tools, we first showcase the performance of the KNN imputation component to \modelnamec, which accepts an arbitrary amount of photometry and infers what the `missing' photometry is. Correctly inferring the missing filters (features) that the property inference models are trained on is critical in the limited information case to avoid causing a biased set of photometric data as inputs to infer galaxy properties. In Figure~\ref{fig:knn_imp}, with our reference sets of filters, we illustrate the performance of the KNN imputer in recovering the original set of photometry points when it only has JWST photometry or HST photometry available. The solid line represents the median offset from the true flux and the shaded regions represent the 16th to 84th percentiles. For the case where only the mock JWST photometry is available, the KNN model typically recovers the shape of the true spectrum well, with a slight bias at the longest wavelengths. Though the imputer typically performs well at short wavelengths, it only underestimates the true flux. However, for the case of HST photometry being available, we note both a larger dispersion and a bias in the FIR in the imputed photometry. Though the bias is not ideal, the high spread is not surprising given that (1) the UV/optical that HST can detect would not offer any likely constraints on the shape of the SED in the IR, and (2) as one observes higher redshift galaxies, the UV will be redshifted out of the observable range of HST, so a large fraction of the sample has even fewer constraints for the imputation. The feature in the HST results at the shortest UV wavelengths is a product of how we store the values of filters that are not detectable at high z and is not physical. The bias may be caused by the spectra in the training set having similar UV/optical shapes but different dust properties. Overall, these results suggest that the imputation step is recovering accurate sets of the photometric points from the constraints it has available.

\section{Model Evaluation}\label{sec:res}

We construct several test cases to evaluate the performance of our ML model relative to the performance of standard SED fitting software (as a reminder, we compare specifically to \textsc{prospector}). Each test case corresponds to different available photometric data and redshift information for \modelname and \textsc{prospector}. We compare the performance in recovering physical properties for the cases where (1) the full photometric and redshift information are available (Section~\ref{subsec:full_phot}), (2) only the JWST photometry is available (Section~\ref{subsec:jwst_phot}), and (3) only the HST photometry is available (Appendix~\ref{sec:HST_phot}) on the test set of galaxies.

For each test case, we evaluate the relative performance on several metrics, listed below. The first 3 metrics listed below correspond to the accuracy of the median predictions of \modelnamec, and the remaining 2 evaluate the confidence interval that \modelname outputs. These are evaluated for each inferred physical property separately. For a given physical property and input instance $i$, $\mu\hat(i)$ represents the prediction of \modelname for the center of the distribution of that property for that input, and $Y(i)$ represents the ground truth of that instance for that property. 

Equation~\ref{eq:nrmse} represents the normalized root mean square error (NRMSE), which reflects how accurate a typical prediction is computed from the offset from the truth squared.
\begin{equation} 
    \label{eq:nrmse}
    NRMSE = \sqrt{\frac{1}{n}\sum_{i=1}^{n}{(\mu\hat (i)-Y(i))^2}}
\end{equation}

Equation~\ref{eq:nmae} is the normalized mean absolute error (NMAE), which reflects how accurate a typical prediction is from the absolute value of the truth. 
\begin{equation}
    \label{eq:nmae}
    NMAE = \frac{1}{n}\sum_{i=1}^{n}{|\mu\hat(i)-Y(i)|}
\end{equation}

Equation~\ref{eq:nbe} represents the normalized bias error (NBE), which reflects how much a typical prediction is biased from the ground truth.
\begin{equation}
    \label{eq:nbe}
    NBE = \frac{1}{n}\sum_{i=1}^{n}{(\mu\hat(i)-Y(i))}
\end{equation}

Equations \ref{eq:ace} and \ref{eq:is} are the average coverage error (ACE) and interval sharpness (IS) respectively. These evaluate whether output uncertainties reflect the true uncertainty in the model. We evaluate these for the 68\% (or $1\sigma$) confidence interval, which means in equations \ref{eq:ace} and $\ref{eq:is}$ $\alpha=1-0.68=0.32$, $l_{\alpha}(i)=0.16.$ and $u_\alpha (i)=0.84$. $L_\alpha(i)$ and $U_\alpha(i)$ correspond to the confidence interval boundaries for the prediction in the space of the predicted variable.
\begin{equation}
    \label{eq:ace}
    ACE_{\alpha} = \frac{1}{n}\sum_{i=1}^{n}{c_\alpha(i)\times100\%-100\times(1-\alpha)\%}
\end{equation}

\begin{equation}
    \label{eq:is}
    c_{\alpha}(i) = \begin{cases}
    1 & Y(i) \in [L_\alpha(i),U_{\alpha}(i)] \\
    0 & Y(i) \notin [L_\alpha(i),U_{\alpha}(i)] \\
\end{cases}
\end{equation}

\begin{equation}
    IS_{\alpha} = \frac{1}{n}\sum_{i=1}^{n}\begin{cases}
    -2\alpha\Delta_{\alpha}(i)-4[l_\alpha(i)-y(i)]& y(i)<l_\alpha(i) \\
    -2\alpha\Delta_{\alpha}(i)-4[y(i)-u_\alpha(i)]& y(i)>u_\alpha(i) \\
    -2\alpha\Delta_{\alpha}(i) & l_\alpha(i)\leq y(i) \leq u_\alpha(i) \\
\end{cases}
\end{equation}      

We compute each of these metrics for both our ML model and \textsc{prospector} outputs for each photometry case and adopt the code from \citep{gilda_mirkwood_2021} to calculate these metrics. For all of these metrics, a value closer to 0 indicates better performance. A summary of our results can be found in Table~\ref{table:all_metric}.

 \begin{table*}[ht]
    \centering
    \begin{tabular}{|c|c|c|c|c|c|c|}
        \hline
        {\bf Property} & {\bf Photometry} & {\bf NRMSE} & {\bf NMAE} & {\bf NBE} & {\bf ACE}& {\bf IS}\\
        \hline
 
        $M_*$ & All & (\textbf{2.22}, 25.36) & (\textbf{0.08}, 0.15) & (\textbf{0.01}, 0.10) & (\textbf{0.11}, -0.41) & (\textbf{-0.785}, -1.41)\\\hline
        $M_*$ & JWST & (\textbf{0.26}, 47.8) & (\textbf{0.08}, 0.22) & (\textbf{0.001}, 0.2) & (\textbf{0.083}, -0.37) & (\textbf{-0.81}, -1.36) \\\hline
        $M_*$ & HST & (\textbf{4.14}, 64.73) & (\textbf{0.11}, 0.31) & (\textbf{-0.2}, 0.26) & (\textbf{0.04}, -0.33) & (\textbf{-0.85}, -1.29) \\\hline

        $M_{\rm d}$ & All & (\textbf{5.54}, 126.24) & (\textbf{0.26}, 0.54) & (\textbf{0.02}, -0.51) & (\textbf{0.05}, -0.63) & (\textbf{-0.86}, -1.71) \\\hline
        $M_{\rm d}$ & JWST & (\textbf{4.25}, 231.86) & (\textbf{0.36}, 1.14) & (\textbf{0.02},-0.94) & (\textbf{-0.11}, -0.40) & (\textbf{-1.04}, -1.39) \\\hline
        $M_{\rm d}$ & HST & (\textbf{10.38}, 68.84) & (\textbf{0.48}, 0.75) & (\textbf{-0.04}, -0.28) & (\textbf{-0.18}, -0.29) & (\textbf{-1.11}, -1.24) \\\hline

        SFR & All & (\textbf{2.59}, 329.47) & (\textbf{0.09}, 1.39) & (\textbf{-0.01}, 1.34) & (\textbf{-0.07}, -0.13) & (\textbf{-0.96}, -1.03) \\\hline
        SFR & JWST & (\textbf{5.42}, 790.20) & (\textbf{0.10}, 3.21) & (\textbf{-0.02}, 3.19) & (\textbf{-0.09}, -0.16) & (\textbf{-0.99}, -1.08) \\\hline
        SFR & HST & (\textbf{21.83}, 1244.48) & (\textbf{0.14}, 5.06) & (\textbf{-0.09}, 5.04) & (\textbf{-0.121}, -0.13) & (\textbf{-1.03}, \textbf{-1.03}) \\\hline

    \end{tabular}

\caption{ \textbf{Our model outperforms traditional SED fitting software on all quantitative performance metrics we use, for all our constructed test cases.} \rm{For all metrics, a value closer to 0 indicates better performance. The first two columns in the table denote the galaxy physical property and the available photometry when evaluating it. For each metric, the first value listed in the tuple references the metric value from \modelname evaluated on the test set, while the second represents the metric from a standard SED fit to the test set data. We bold the lower of the two values in each tuple. In order, the metrics are normalized root mean square error (NRMSE), normalized mean absolute error (NMAE), normalized bias error (NBE), absolute coverage error (ACE), and interval sharpness (IS). For all of these test metrics, \modelname performs as well as or better than the standard SED fitting software.}}
\label{table:all_metric}
\end{table*}

\subsection{Full photometry/ known redshift case}\label{subsec:full_phot}

We first compare the performance of \modelname and traditional SED fitting in correctly inferring the physical properties of our simulated galaxies from their spectra with access to the true redshift and all the possible photometric filter information and assuming an SNR of 10. This test represents the best-case scenario, with the maximum amount of constraining information. 

\begin{figure*}[htp]
    \epsscale{1.25}
    \plotone{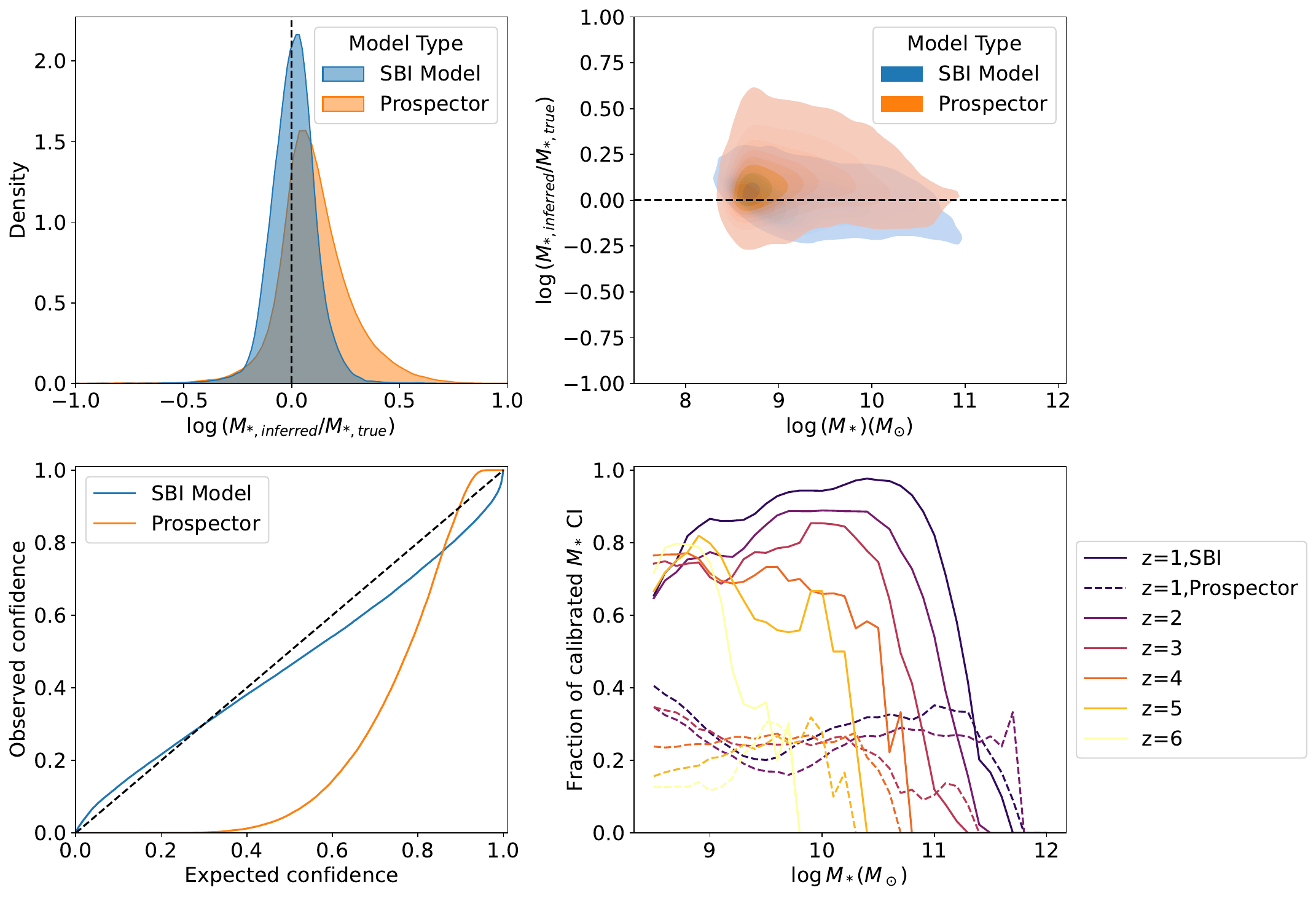}
    \caption{\textbf{Comparison of \modelname and traditional SED fitting in recovering galaxy stellar mass and the output confidence intervals when all training filters and redshift are available as inputs.} Our model recovers galaxy stellar mass better than traditional SED fitting and is also largely returning appropriate uncertainties. Top left: simple 1D comparison for accuracy for stellar mass with all trained filters available and redshift known. Top right: comparison of accuracy as a function of galaxy stellar mass. Bottom left: Observed vs expected uncertainty calibration plot. The percentiles of model predictions and uncertainties are compared to the offset between the prediction and the ground truth value. `Perfectly calibrated' uncertainties will lie on the dashed 1-to-1 line. Bottom right: Comparison between the fraction of confidence intervals in a given stellar mass and redshift bin where that contain the ground truth stellar mass. Solid lines represent \modelname and dashed lines represent traditional SED fitting. Any individual line represents the fraction of output 68\% CIs for a given ground truth stellar mass by either \modelname or traditional SED fitting where the ground truth stellar mass is within the inferred CI.}
    \label{fig:mstar_full}
\end{figure*}

\begin{figure*}[t]
    \epsscale{1.25}
    \plotone{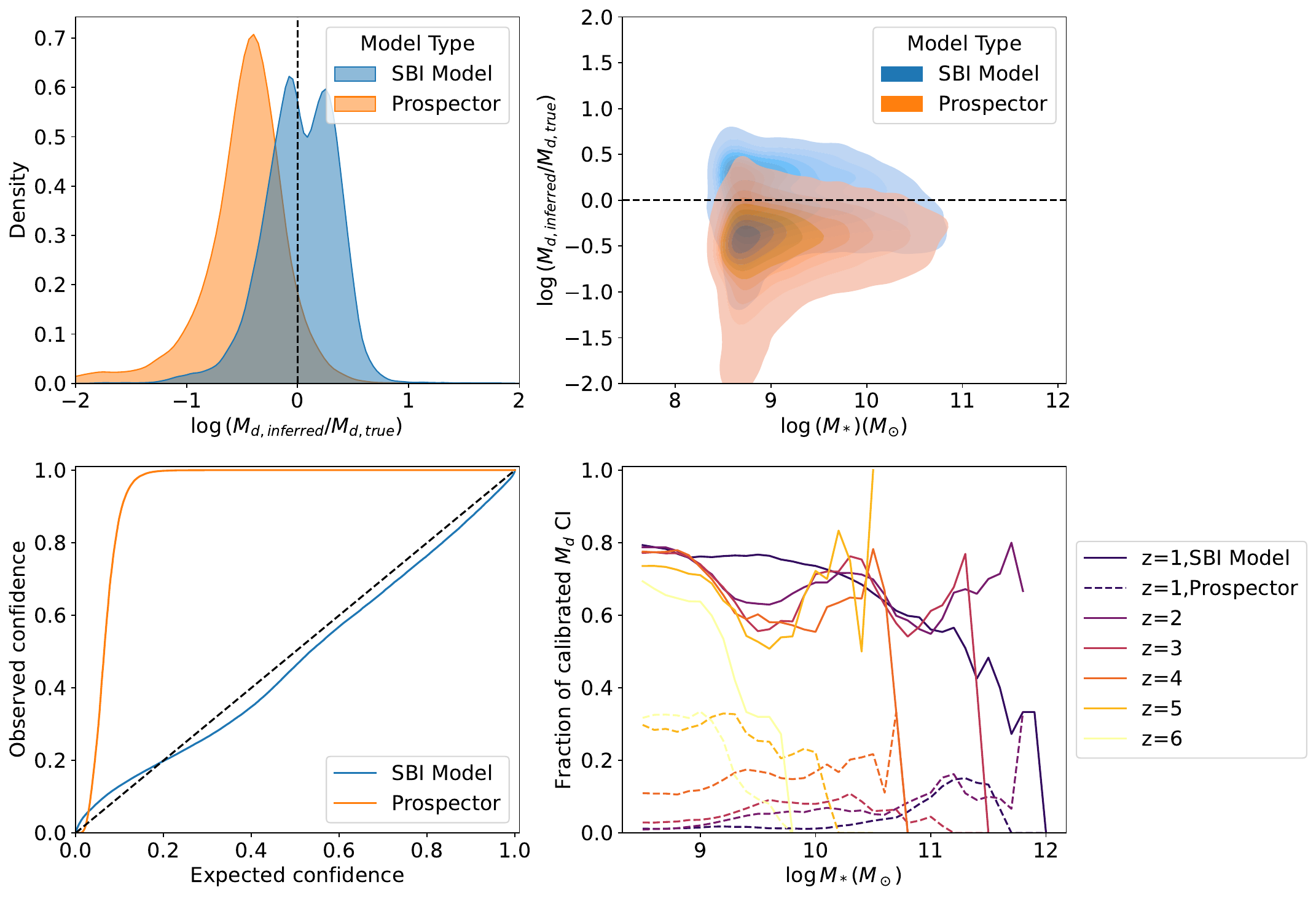}
    \caption{\textbf{Similar to Figure~\ref{fig:mstar_full}, but for galaxy dust mass inference with all redshift and photometric data available as constraining information.} Top left: simple 1D accuracy comparison between \modelname and traditional SED fitting for recovering galaxy dust mass. \modelname has a double-peaked distribution tied to the two galaxy formation models that make up our training set, while traditional SED fitting underpredicts dust mass by $\sim0.5$ dex. Top right: The lowest mass galaxies tend to have the largest spread in the dust mass accuracy. Neither \modelname nor traditional SED fitting seems to have a strong mass dependence on the bias. Bottom left: Observed vs expected uncertainty calibration plot for dust mass inference. \modelname has well-calibrated errors. Bottom right: Comparison between the fraction of confidence intervals in a given stellar mass and redshift bin where that contain the ground truth stellar dust mass. The reported uncertainties of \modelname contain the true dust mass significantly more often than traditional SED fitting.}
    \label{fig:md_full}
\end{figure*}

\begin{figure*}[t]
    \epsscale{1.25}
    \plotone{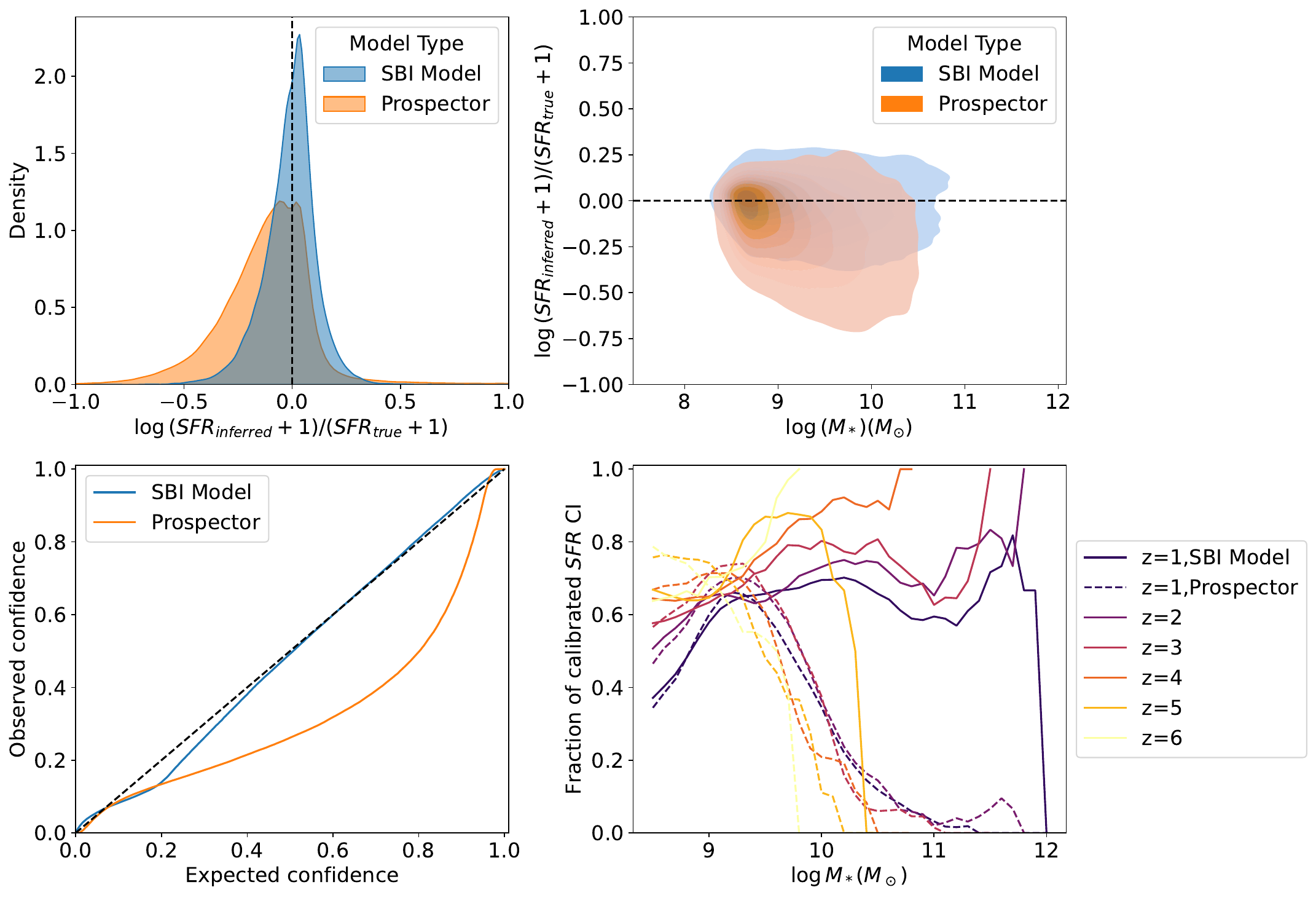}
    \caption{\textbf{Similar to Figure~\ref{fig:mstar_full}, but for galaxy SFR inference with all redshift and photometric information available as model inputs.}}
    \label{fig:sfr_full}
\end{figure*}

\modelname outperforms standard SED fitting software in recovering the galaxy stellar mass.  To demonstrate this explicitly, in Figure~\ref{fig:mstar_full} we present a comparison between the performance of \modelname and \textsc{prospector} in recovering the true stellar mass of the galaxy in this ideal scenario. We see in the top left panel that traditional SED fitting exhibits a minor bias in its predicted stellar masses relative to our ML model and has a broader dispersion. The bias is more clearly seen in the top right hand panel of Figure~\ref{fig:mstar_full}, where we compare the offset between the inferred and true stellar mass as a function of galaxy stellar mass. We see from the right side of Figure~\ref{fig:mstar_full}, that standard SED fitting software overall is not strongly biased, but exhibits a large spread in performance across a range of stellar masses. We note that there is a slight bias for \modelname to predict the mass of the low mass galaxies in the test set high and the mass of high mass galaxies slightly low; this behavior is a common feature of ML models. The scarcity of the most massive galaxies in our training set also likely contributes to the bias at the high mass end.

In the lower panels of Figure~\ref{fig:mstar_full}, we evaluate the performance of the ML model and \textsc{prospector}'s output confidence intervals (CIs). In the left panel, we present a typical error calibration plot often used in ML studies. This plot is constructed by evaluating the cumulative distribution percentile from the output posterior of the actual ground truth for each input instance. These are then ordered, and if all the uncertainties are perfectly calibrated relative to the ground truth, then the expected confidence percentile should equal the `observed' confidence percentile. The 1:1 line that represents this concept of perfectly calibrated uncertainties; being above the dashed line represents overconfidence, and being below the dashed line represents underconfidence.  Though this metric is commonly used in ML studies, it is intended to evaluate whether the uncertainty associated with a quantity reflects what the true uncertainty is, so it fair to apply it to the uncertainties derived from the SED fitting posteriors. The relative offsets between \modelname \ and \textsc{prospector} suggest that the ML model may have slightly better calibrated errors than standard SED fitting techniques. 
The lower right panel illustrates this information in further detail by binning the test galaxies into bins of redshift and stellar mass and calculating the fraction of confidence intervals (CIs) in that bin where the true value lies within the stated uncertainties; for example, this means that a calibrated fraction of 0.8 at $z=6$ and a stellar mass of $10^9 M_\odot$ means that 80\% of the galaxies at $10^9 M_\odot$ and $z=6$ have a 68\% confidence interval between the 16th and 84th percentiles of the output posterior that includes the ground truth. Our ML model in general is best-calibrated for the lower mass galaxies in the sample; given that these galaxies represent the vast majority of the training set, this is not surprising. Similarly, our uncertainties are typically worse for the most massive galaxies at any given redshift given their scarcity. However, the \textsc{prospector} CIs in any given stellar mass and redshift bin are typically not more than $\sim20\%$ calibrated and are only comparable to \modelname for the highest mass galaxies at a given redshift. We see that the not only is \modelname typically more accurate than traditional SED fitting in inferring $M_*$ in this ideal case it was trained on, but it also is more appropriately confident in its predictions.

In Figure~\ref{fig:md_full}, we perform a similar comparison with the predicted dust mass of galaxies from \modelnamec. \modelname should theoretically have a significant advantage for this property, given the diversity of star-dust geometry in the 3D radiative transfer used to generate our training and testing set rather than assuming a uniform dust screen, as is typically done in standard SED fitting software. Traditional SED fitting typically underpredicts the galaxy dust masses by $\sim0.5$ dex, but the distribution does include correctly recovering the initial mass. Our dust mass ML model has a bimodal feature in its performance but does not exhibit the bias present in standard SED fitting software. The bimodality present in the \modelname results can likely be attributed to the bimodal distributions of dust mass present in the \textsc{simba} and \textsc{IllustrisTNG} simulations. The success also holds when examining the accuracy in recovering dust mass as a function of stellar mass in the upper right panel; the ML model's performance does not significantly change as a function of stellar mass and \textsc{prospector} has a systematically offset distribution which does not have a strong correlation with the stellar mass. The behavior of \textsc{prospector} is not surprising - we have set it to assume a uniform dust screen, and our current radiative transfer modeling assumes that the dust attenuation is in diffuse dust throughout the galaxy mixed with the stars in the star-dust geometry in the simulated galaxies. Because of star-dust geometry in radiative transfer, it is possible that not all the starlight will be attenuated by all the dust, contrary to the assumption of the uniform dust screen. Therefore, the attenuation that \textsc{prospector} derives would be expected to underestimate the dust mass of the system. In theory, \textsc{prospector} can have some flexibility in the fit attenuation law, though we note our initial testing with this did not yield significantly different inferred dust masses.

In the lower panels of Figure~\ref{fig:md_full}, we assess the uncertainties in the dust inference. We once again see that \modelname is reasonably well-calibrated, though it is slightly underconfident based on the calibration plots. However, the error calibration from standard SED fitting software is relatively poor for dust mass. Investigating the poor calibration in more detail, we see that the the CIs for dust from the ML model generally perform similarly to those for for $M_*$ but do not have a significant reduction in performance for the most massive galaxies. The 68\% CIs inferred from \textsc{prospector} fitting typically do not reflect the offset from the true value. It is possible that the dust attenuation laws may differ significantly from the model extinction law in these cases, in which case the assumption in our \textsc{prospector} priors may be too restrictive and therefore it may become overconfident in the dust properties in the highly constrained spectrum we are providing.

Finally, in Figure~\ref{fig:sfr_full}, we present the inference of SFR for the ideal scenario (i.e. the all photometry and known redshift case that we have been studying this entire subsection). \modelname and \textsc{prospector} both have their distribution peaks at little to no offset from the truth, but \textsc{prospector} has a much broader spread from the true values than the predictions from \modelnamec. We note that as a function of stellar mass, \modelname is rather successfully predicting the true SFR, whereas traditional SED fitting typically underestimates the SFR in the more massive galaxies in the test set.

We next evaluate our model uncertainties in recovering the SFR. In the lower panels of Figure~\ref{fig:sfr_full}, we find several interesting behaviors. We note that again the uncertainties from our ML model are reasonably well-calibrated; the lower deviation from the `perfectly calibrated' line for \textsc{prospector} suggests a greater proportion of reasonable uncertainties. Based on the second panel, we can attribute the apparent calibration to the fact that for the galaxies in the training set at lowest masses, the \textsc{prospector} CIs are typically be appropriately sized, though they do not reflect the true uncertainty at higher stellar masses, which are rarer in the sample. \modelname is comparable to \textsc{prospector} in its inferred CIs at lower stellar masses, though it continues producing properly sized CIs at higher stellar masses.

\subsection{JWST photometry/unknown redshift}\label{subsec:jwst_phot}

\begin{figure*}[b]
    \epsscale{0.75}
    \plotone{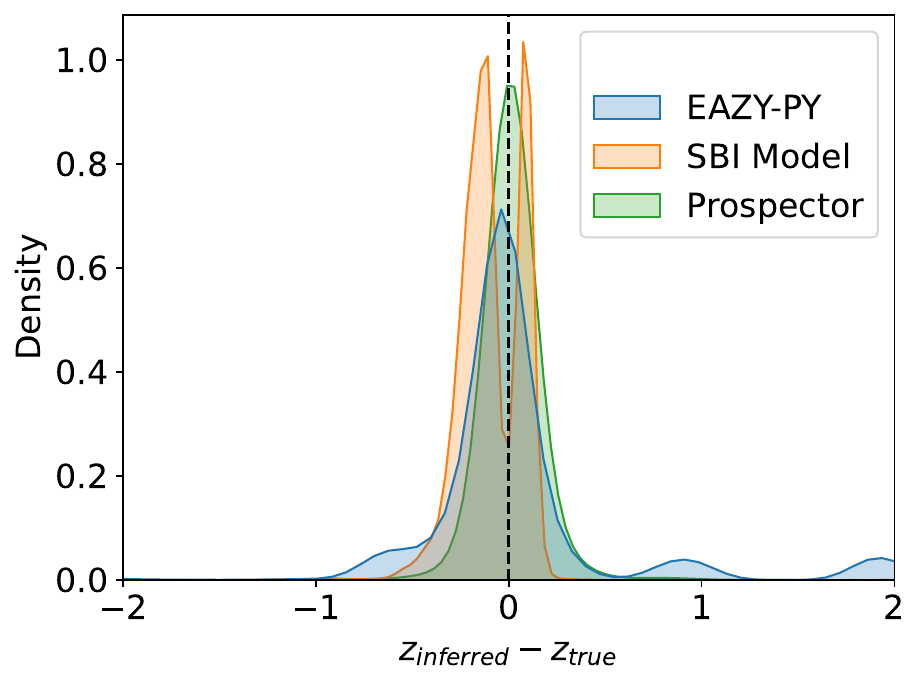}
    \caption{\textbf{Comparison of the performance of \modelname with EAZY-PY and \textsc{prospector} for recovering the true redshift with only JWST MIRI and NIRCam photometry available.} }
    \label{fig:z_jwst}
\end{figure*}

As we have just discussed (\S~\ref{subsec:full_phot}), \modelname is generally successful in reproducing important galaxy properties and outperforms standard SED-fitting software when considering the ideal/limiting case where it has access to all the photometry it is trained on and the redshift is precisely known. We now turn to a more realistic case where the observations do not have access to all wavelengths the model is trained on or the redshift.  Here, we require the use of the KNN imputer and our photo-z module to infer properties. As an example, we first limit the `observations' to only those filters corresponding to what JWST would observe with the NIRCam and MIRI wide filters and assume the redshift is not known. We therefore evaluate the performance of the inferred redshift \modelname derives in addition to the other known physical properties. We remind the reader that in this configuration, we also allow the redshift to be free for \textsc{prospector} for fairer comparisons with uncertainties and benchmark our photo-z against the common photo-z software \textsc{eazy-py} \citep{brammer_eazy_2008}. In Appendix~\ref{sec:HST_phot}, we repeat this exercise for the case where only HST photometry is observed.

\begin{figure*}[b]
    \epsscale{1.25}
    \plotone{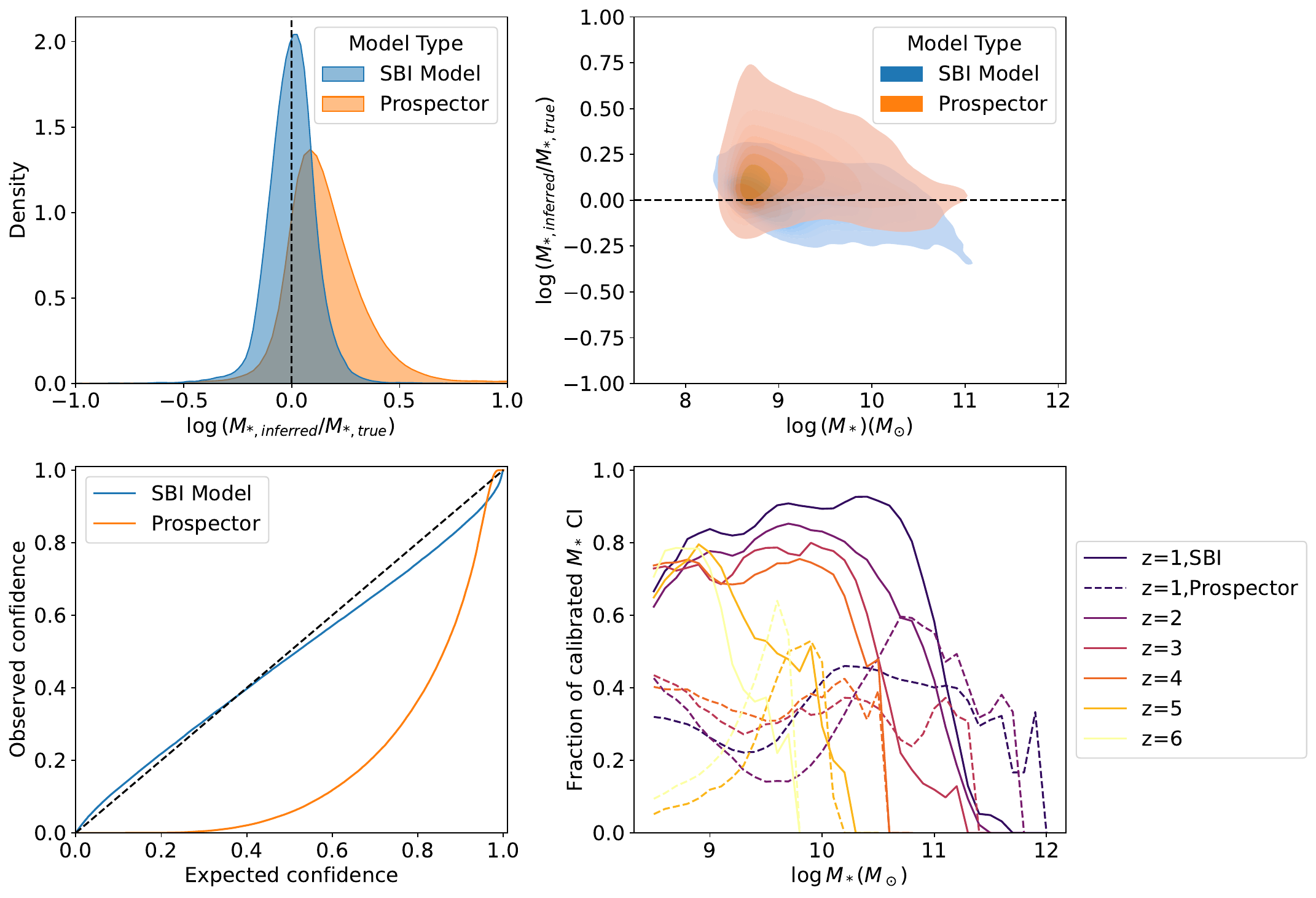}
    \caption{\textbf{Similar to Figure~\ref{fig:mstar_full}, but now with only JWST photometry available and no redshift constraints as model inputs.} }
    \label{fig:mstar_jwst}
\end{figure*}

\begin{figure*}
    \epsscale{1.25}
    \plotone{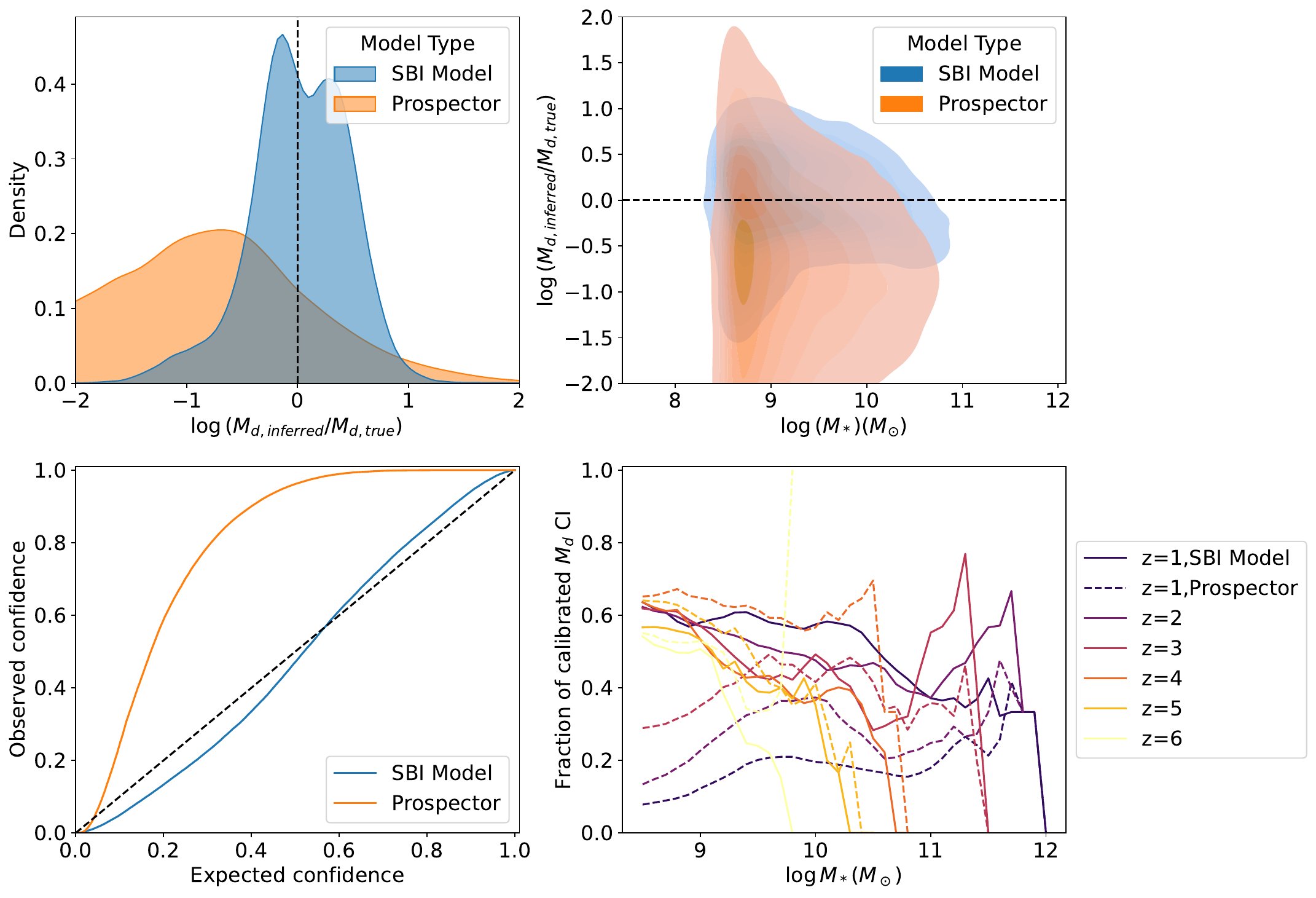}
    \caption{\textbf{Similar to Figure~\ref{fig:mstar_jwst}, but for galaxy dust mass inference with JWST photometry only and no redshift constraints as model inputs.}}
    \label{fig:md_jwst}
\end{figure*}

\begin{figure*}
    \epsscale{1.25}
    \plotone{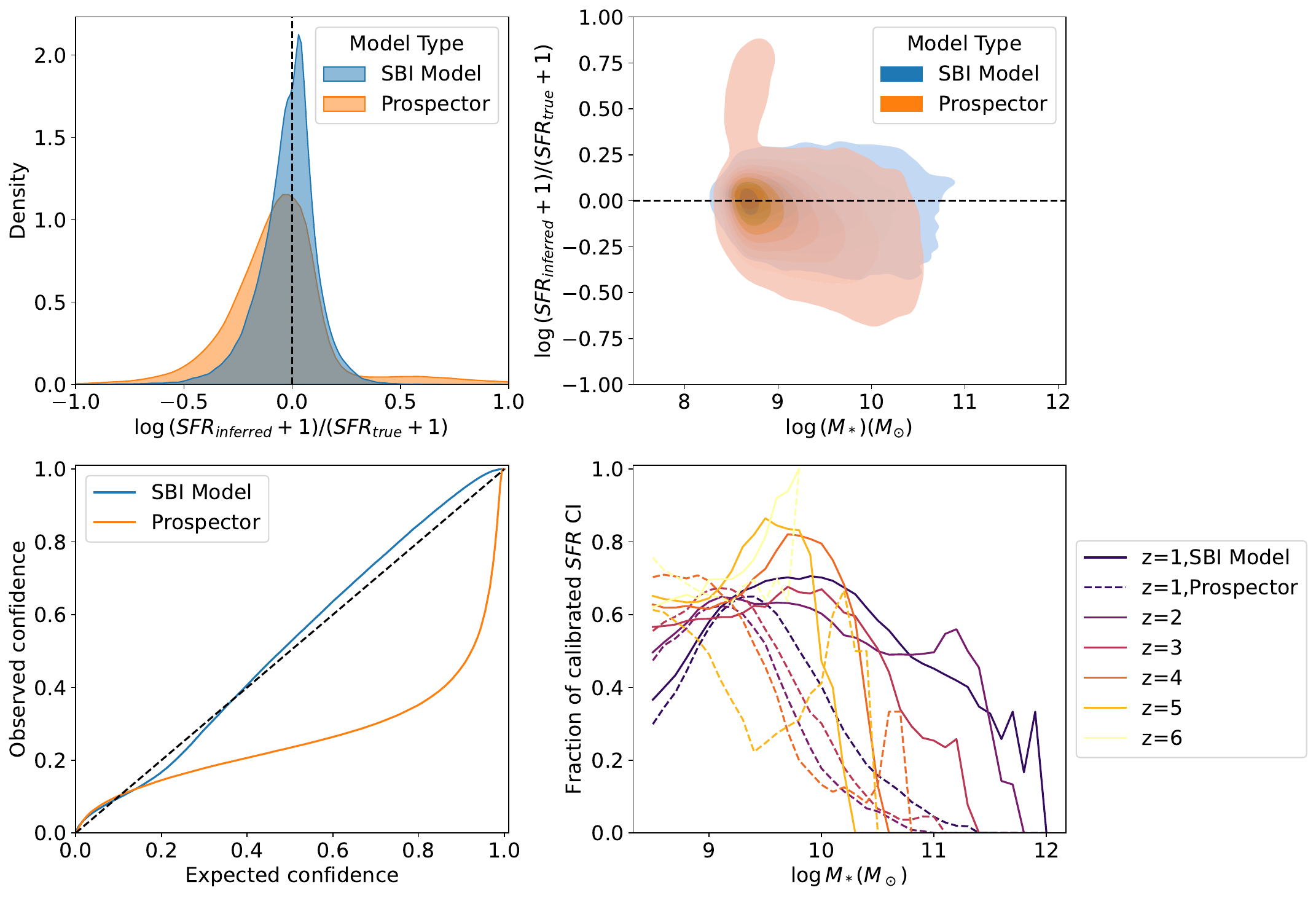}
    \caption{\textbf{Similar to Figure~\ref{fig:mstar_jwst}, but for galaxy SFR inference with JWST photometry only and no redshift information as model inputs.}}
    \label{fig:sfr_jwst}
\end{figure*}
In Figure~\ref{fig:z_jwst}, we present the results of the aforementioned photo-z tests. We find that with JWST photometry available, \modelname typically either slightly overpredicts or underpredicts the true redshift. However, \textsc{eazy-py} exhibits tails of outliers of $|z_{inferred}-z_{true}|>2$ that are not present in our model. We also include the performance of \textsc{prospector} in recovering the redshift as well for comparison. \textsc{prospector} exhibits slightly larger tails in offset for redshift but overall is more concentrated around the true redshift and does not have a subset of catastrophic outliers.

In Figures~\ref{fig:mstar_jwst}, \ref{fig:md_jwst}, and \ref{fig:sfr_jwst}, we present the results for running our SED modeling tool as compared to traditional SED fitting with the test sample being `observed' with on JWST photometry. We generally note some degradation in performance now that the full information is not available, but overall the ML models continue to infer galaxy properties successfully. Notably, the spread in dust mass predictions relative to the ground truth for \modelname has increased. Also, the performance of the CIs for the ML model has noticeably decreased; therefore, the uncertainties \modelname is deriving are not fully scaling with the additional uncertainty being added from missing photometric and redshift information and therefore that the MC imputation step is not fully reflecting the uncertainties that it is introducing (see Section \ref{sec:res}).

For stellar mass inference in Figure~\ref{fig:mstar_jwst}, the distribution of offsets from the true value by \modelname has slightly increased. However, the bias in \textsc{prospector} has increased and the distribution has widened mildly. For the CIs, there is a noticeable reduction in performance for \modelnamec, where the character stellar mass threshold at each redshift where the uncertainties are no longer reliable has shifted to lower mass. The \textsc{prospector} $M_*$ CIs are more properly sized than before as the availability of photometry has decreased and redshift is a free parameter, but typically caps at only $\sim40\%$ of the derived uncertainties in any stellar mass bin being reasonable.

For the dust mass predictions in Figure~\ref{fig:md_jwst}, the bimodal feature of the distribution has not changed, but the spread has increased substantially from the initial results as the additional limitation of unknown redshifts and limited photometry is added to the system. There also is a noticeable performance decrease in our CIs for dust masses in with JWST photometry; again, the poorer performance implies the CIs \modelname is deriving are not fully scaling with the additional uncertainty from missing photometric constraints as they should. However, the distribution of inferred dust mass for standard SED fitting is much broader without the constraints imposed in the full photometry case, with some estimates being off by more than 2 dex. However, \textsc{prospector} also has its stated uncertainties increase, especially at high redshift, with JWST photometry only, so the \textsc{prospector} uncertainties tend to reflect the true uncertainty more often than in the `complete photometry' test case. 

The SFR performance (Figure~\ref{fig:sfr_jwst}) changes minimally for both \modelname and traditional SED fitting with only JWST photometry available. However, the main consequence is a decrease in CI performance for the ML model for the most massive galaxies at each redshift. It is less surprising that the imputation step is not appropriately adding uncertainties for the most massive galaxies since they are the least represented in the training set. The \textsc{prospector} CIs are also similar to their performance in the full photometry case.

\section{Model Exploration}\label{sec:model_ex}
Having compared our fiducial ML models and \textsc{prospector}, we now investigate each of the components of \modelname in to understand the source of the models output in more detail. In this section we (1) produce and discuss Shapley (SHAP) value feature importance plots to understand which photometry is most important to successfully compute a given physical property of galaxies, (2) investigate how each source of uncertainty in our modeling contributes to the final confidence interval inferred, (3) evaluate the regimes in which \modelname performs the most poorly in recovering stellar mass, and finally (4) investigate the epistemic uncertainties associated with using cosmological simulations to construct an SED modeling software by (a) training a new model solely on \textsc{simba} data and testing on \textsc{IllustrisTNG} data and (b) evaluating our fiducial model against galaxies/spectra produced from a cosmological simulation that whose properties the model is not trained to reproduce.

\subsection{SHAP results}\label{subsec:shap}
\begin{figure*}[t]
    \epsscale{0.75}
    \plotone{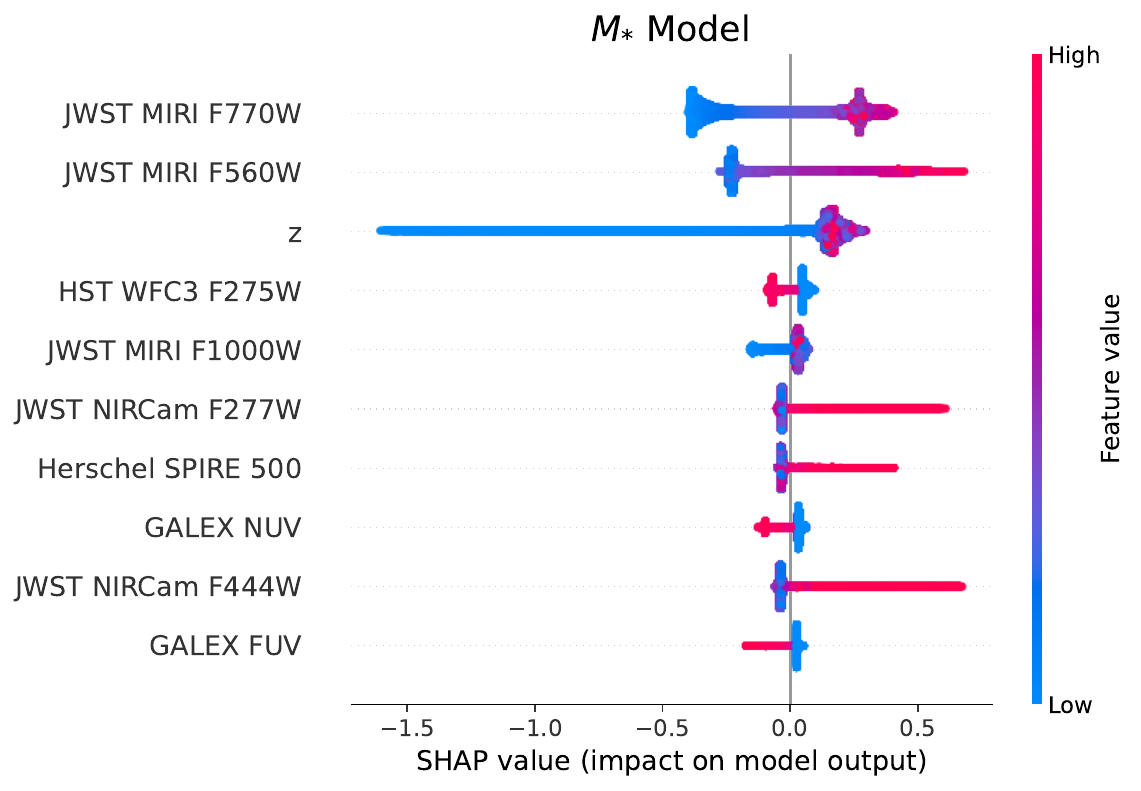}
    \caption{\textbf{A SHAP value plot for the \textsc{NGBoost} stellar mass inference model in \modelname indicates that the rest-frame K band is primarily what \modelname uses to predict stellar mass.} Input features are ordered from highest to lowest for how much the ML model relies on the feature to recover the stellar mass. High absolute SHAP values mean that the property is more important to the output prediction. The shading for any individual feature reflects whether the SHAP values of that normalization correspond to the higher or lower values of that feature in the sample.}
    \label{fig:shap_sm}
\end{figure*}
\begin{figure*}[bt]
    \epsscale{0.75}
    \plotone{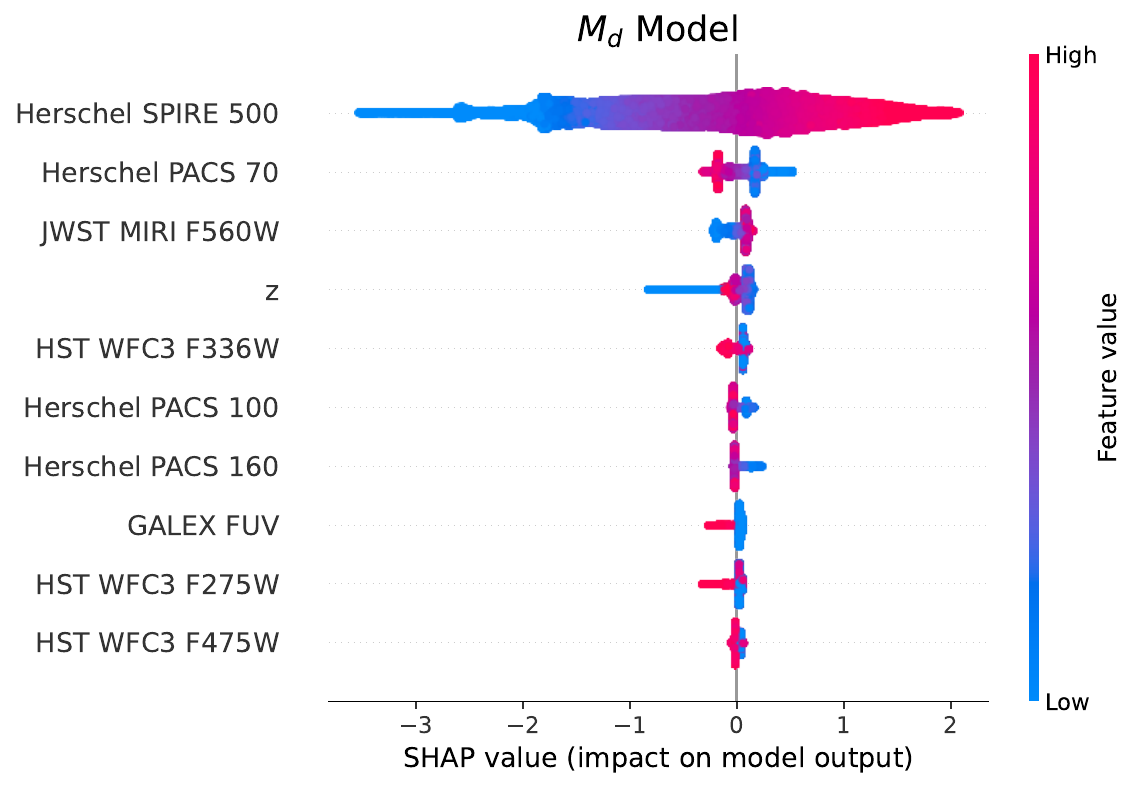}
    \caption{ \textbf{Similar to Figure~\ref{fig:shap_sm} -  a SHAP analysis for the dust mass portion of \modelname indicates that the far-IR is what the model uses to recover galaxy dust masses. }}
    \label{fig:shap_dm}
\end{figure*}
We begin dissecting \modelname by investigating what photometric filters (and redshift information) are the most important to each trained model to infer galaxy properties by using \cite{shapley_stochastic_1953} (hereafter SHAP) value plots. SHAP values can be used to evaluate how much each `player' (feature in this case) contributes to the `prize' (prediction) for a model. SHAP values also account for features not being independent, which is essential to evaluate \modelname where photometry of the same spectrum will naturally not be independent. In these plots, the distribution of SHAP values for a particular input feature from the test set are represented in each row. Higher absolute SHAP values indicate that the feature was more important to the output prediction. For a given feature, the red and blue colors represent that the SHAP value is from high and low range of the distribution of feature values. Features are in descending order in these figures such that the top feature is the one with the highest typical SHAP value (most important to the model predictions), and the typical SHAP value decreases from that.

\begin{figure*}
    \epsscale{0.75}
    \plotone{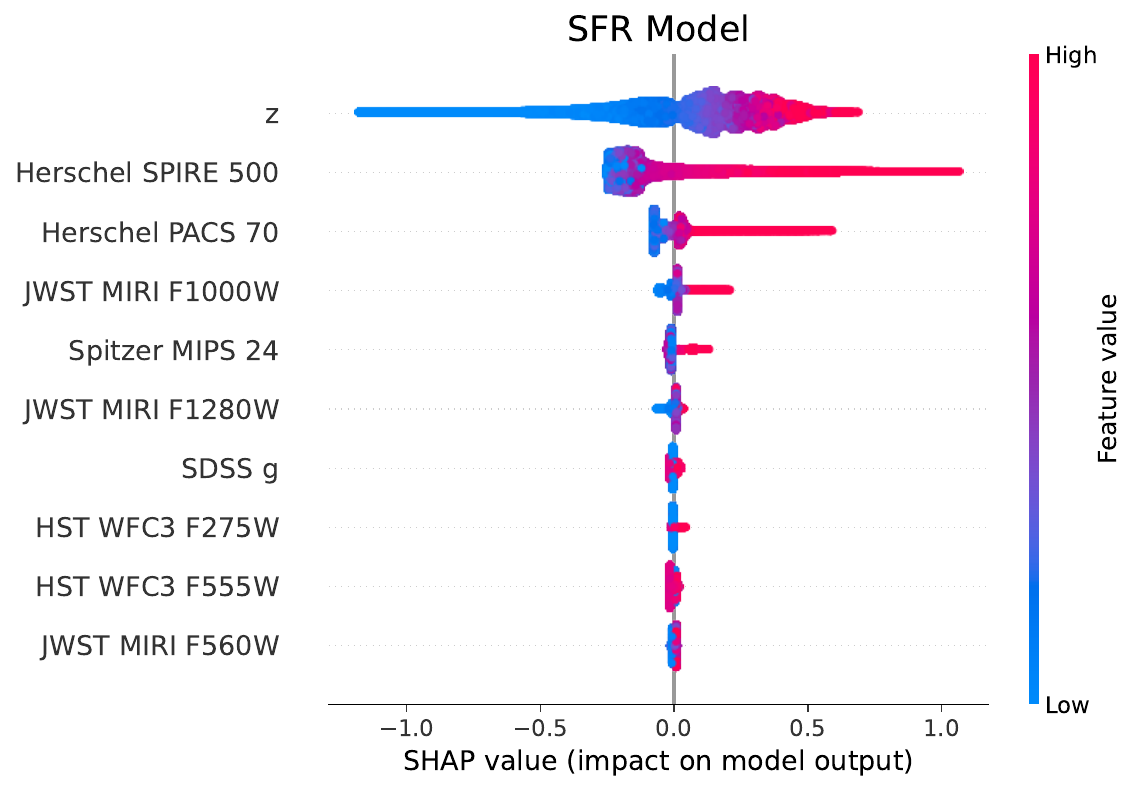}
    \caption{ \textbf{Similar to Figure~\ref{fig:shap_sm} - a SHAP analysis for the SFR portion of \modelname indicates that it uses a combination of redshift and far-IR bands to recover star formation rates. }}
    \label{fig:shap_sfr}
\end{figure*}

In Figures~\ref{fig:shap_sm} - \ref{fig:shap_sfr}, we present SHAP plots for the  the filters, in addition to redshift, that each {\sc NGBoost} model relies on the most to infer each respective property. For $M_*$ in Figure~\ref{fig:shap_sm}, the 1-10 $\mu m$ range is most important for \modelname to be able to infer stellar mass in addition to knowledge of the redshift. Our SHAP analysis therefore concludes, naturally, that the ML model responsible for inferring galaxy stellar mass is relying most heavily on the rest-frame $K$ band. To infer the dust mass (Figure~\ref{fig:shap_dm}), the model relies most heavily on the highest wavelength Herschel filter at $\sim$500 $\mu$m and in general the IR. There is also some mild dependence on the mid-IR and far-UV. The far-IR continuum being the most important to successfully recover galaxy dust masses is consistent with the work of \cite{draine_dust_2007}.

Finally, to infer the SFR, Figure~\ref{fig:shap_sfr} suggests the model references mainly the redshift and far-IR information. The normalization of the SFMS is highly dependent on redshift, and the majority of the sample is star-forming galaxies. Based on the SHAP plot for stellar mass, it is also clear that the model relies on redshift to correctly place the normalization of the spectrum. Given these two simple statements, it is not surprising that the model uses redshift as its primary feature to recover SFR. Additionally, the significant dependence on the far-IR follows common scaling relations between and $\rm L_{IR}$ and SFR such as \cite{kennicutt_star_1998}. It is likely that the reliance on the dust continuum emission rather than the UV for the ML model to infer the galaxy SFR is because a galaxy is less likely to be optically thick to far-IR photons, and therefore any scaling between $\rm L_{IR}$ and SFR will be simpler. Furthermore, much of the UV light associated with star formation in the Universe is expected to be obscured by dust and remitted in the IR (e.g., \citealt{whitaker_constant_2017}). \cite{zimmerman_tracing_2024} also model that most of the UV is obscured by dust in the fiducial \textsc{simba} 25 Mpc/h volume, so our sample, which includes \textsc{simba} volumes and \textsc{IllustrisTNG} galaxies that we assume to have significantly more dust than analogous \textsc{simba} galaxies, would likely have this trait.

\subsection{Understanding model CIs}\label{subsec:ci_invest}

In Section \ref{sec:res}, we found that \modelname CIs more often underestimated true uncertainty for the test cases where it had limited access to photometric bands. In this section, we now break down \modelname CIs to understand which steps are contributing the most to the width of the posterior for stellar mass inference. 

\begin{figure*}[t]
    \epsscale{0.85}
    \plotone{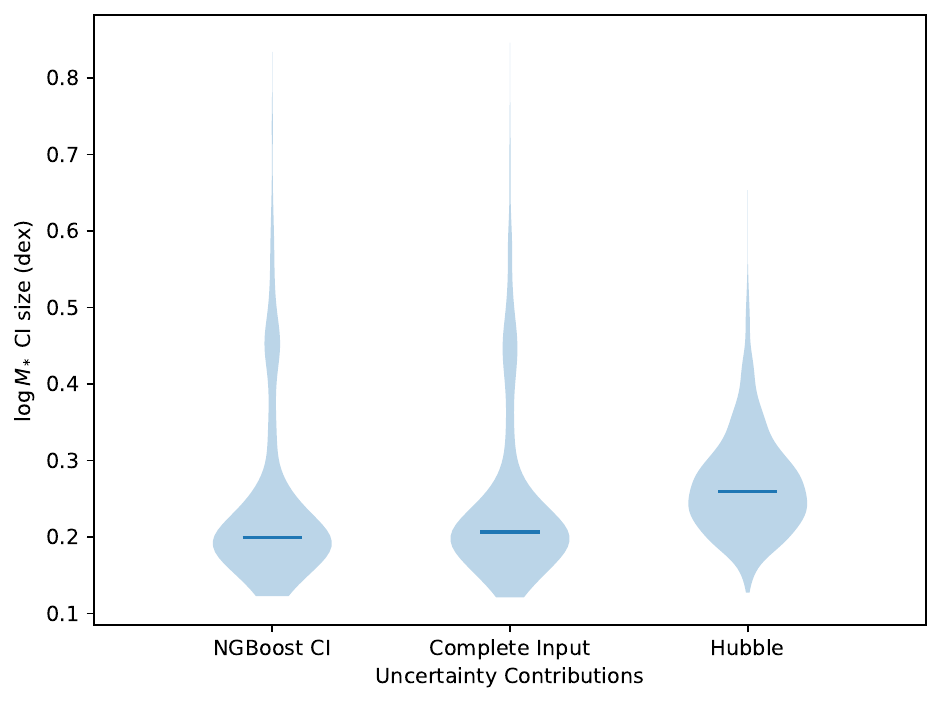}
    \caption{ \textbf{The NGBoost model output standard deviation dominates the contributions to the output posterior.} We present a set of violin plots explaining the breakdown of uncertainty source contributions to \modelname's final CI output for the stellar mass inference. From left to right, another source of uncertainty is added to the test to show the contribution of each of the components of \modelname to the size of the final posterior. The \textsc{ngboost} model typically produces uncertainties of $\sim0.2$ dex in stellar mass; adding in the MC photometry step does little; finally, limiting inputs to only HST photometry (and therefore requiring the use of the imputer) shifts the typical CI size slightly to $\sim0.25$ dex.}
    \label{fig:unc_break}
\end{figure*}

Overall, there are three major contributions to the total uncertainty for a typical observation. These include the uncertainty in imputation/missing photometry, the redshift uncertainty, and the individual {\sc NGBoost} uncertainties. We remind the reader that {\sc NGBoost} models output predict Gaussian distributions with a mean and standard deviation that we draw from on each iteration.  In Figure~\ref{fig:unc_break}, we visualize the typical contribution to output confidence intervals from each of these contributions as violin plots with increasing sources of uncertainty. The leftmost violin reflects when z and the photometry are all known and the only uncertainty is from the {\sc NGBoost} output predicted Gaussian distributions' standard deviations. The middle violin includes the Monte Carlo photometric uncertainty with a SNR of 10 propagated into the final posterior (i.e. the limiting best-case scenario from Section~\ref{subsec:full_phot}), and the final violin mimics a typical input for the case where only HST photometry is available and the redshift is unknown (Section~\ref{sec:HST_phot}). The latter case is to demonstrate the uncertainty in the model results with limited photometry and unknown redshifts. The major takeaway from this test is that the output {\sc NGBoost} CIs dominate over the uncertainty from randomly sampling the photometry uncertainty and that at minimum the uncertainty on any given inference should be $\sim0.1$ dex, and typically should be $\sim0.2$ dex. However, adding in the uncertainty from the imputation step to resolve the missing photometry in combination with the MC sampling does cause the typical CI size to increase from $\sim$0.2 dex to $\sim$0.25 dex. This is not a significant increase, but it is clear that the whole distribution has systematically shifted to higher CI sizes. However, this minor shift explains 
the issues we have noted in Section~\ref{subsec:jwst_phot} and Appendix \ref{sec:HST_phot} that \modelname has fewer inferred CIs that contain the ground truth when it has access to limited input photometry as opposed to the ideal case; the MC photometry and imputation step is not adding sufficient uncertainty to the results to reflect the loss of information, except in a minority of cases.

\subsection{What causes \modelname to perform poorly on certain galaxies?}\label{subsec:bad_gals}

\begin{figure*}
    \plotone{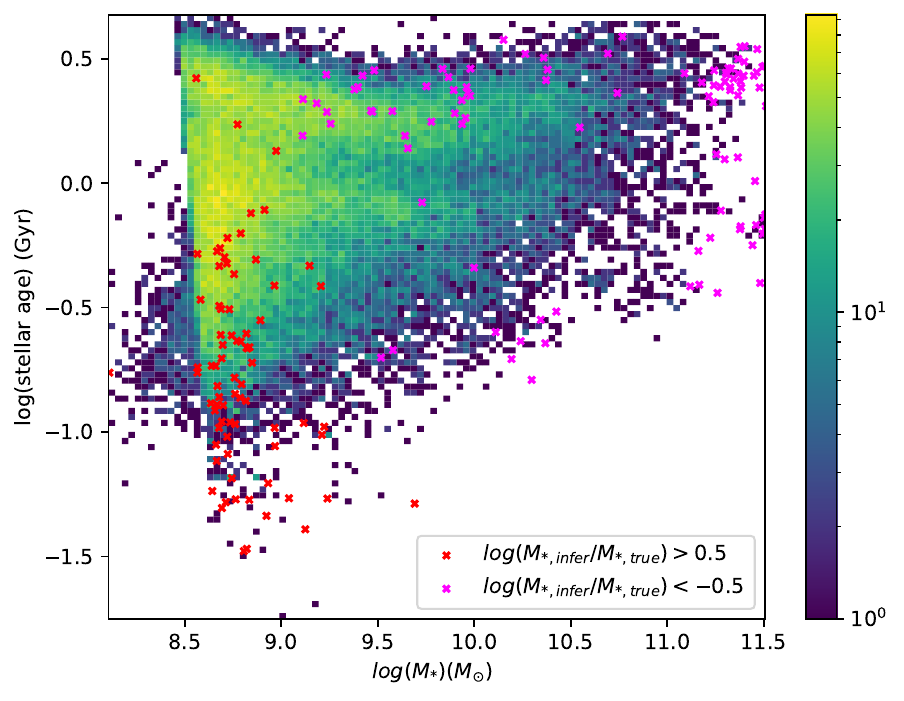}
    \caption{\textbf{\modelname performs the worst in recovering stellar mass on galaxies with stellar ages and masses in areas of parameter space that are sparsely populated in the training set.} Red crosses indicate that \modelname overestimates the true stellar mass and magenta crosses indicate that it underestimates the true stellar mass by at least 0.5 dex. The histogram reflects the properties of the galaxies that make up the training set. It is clear that the galaxies \modelname struggles with the most extreme values of stellar age. Extremely old or young ages implies a different relation between brightness and mass for these galaxies.}
    \label{fig:outlier}
\end{figure*}

In Figure~\ref{fig:outlier}, we investigate the performance of the worst outliers in the inference of galaxy stellar mass in the case where \modelname has access to all the available photometry and the redshift as an example to understand what parts of galaxy parameter space, if any, \modelname struggles to model. The crosses represent inferred stellar masses by \modelname that are at least 0.5 dex offset from the ground truth, while the histogram represents the distribution of galaxies in the training set. In theory, there is little reason for \modelname to perform poorly in these cases with access to information across the SEDs. In Figure~\ref{fig:outlier}, the galaxies that \modelname models poorly are largely the most and least massive galaxies. This behavior is not surprising, as it is common for ML algorithms to under(over)estimate the the largest (smallest) values in any given training set. Additionally, the most massive galaxies are sparsely represented in the training set given both their rarity and the small CAMELS 1P volumes, so these would be challenging for the model to evaluate with few representatives to learn from. Furthermore, we also plot the average stellar age of these galaxies against the training set. Galaxies whose inferred stellar mass is too high are largely the ones with the youngest stars, while galaxies with underpredicted stellar mass tend to be the ones with the oldest stars. This trait makes physical sense because a galaxy whose stellar population is dominated by young (old) stars will appear brighter (dimmer) than otherwise expected for a galaxy with average stellar ages at any given redshift. Overall, given that these galaxies are near the bounds for stellar masses in the training set, rare and/or do not represent the typical mass-to-light relationships for their mass in the training set, it makes sense that \modelname struggles to reproduce their properties accurately.

\subsection{Understanding epistemic uncertainties in our model}\label{subsec:train_test}

\begin{figure*}
    \epsscale{1.2}
    \plotone{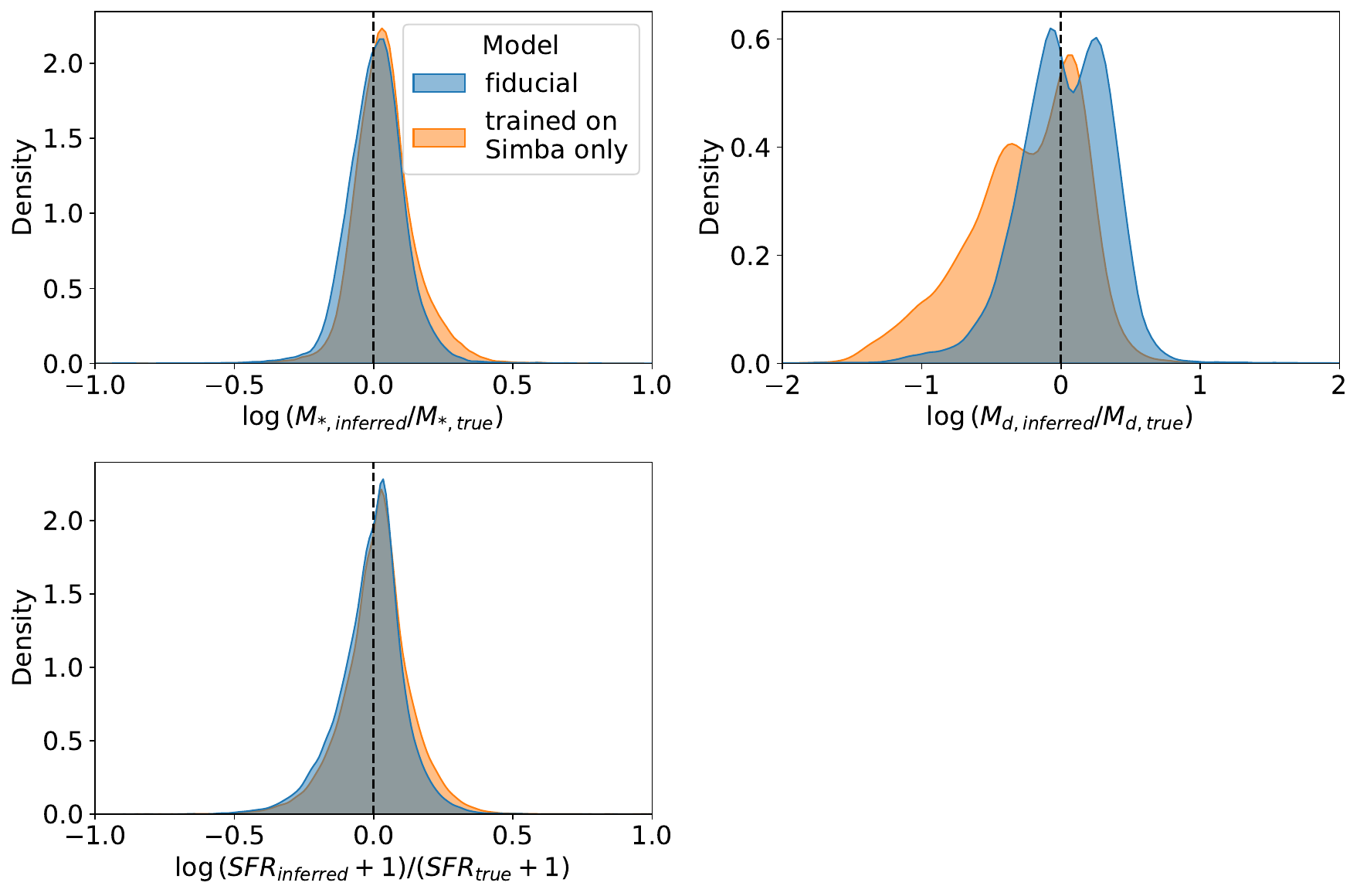}
    \caption{\textbf{Comparison of performance of fiducial models in \modelname on the testing set against models only trained on \textsc{Simba} galaxies and tested on \textsc{IllustrisTNG} galaxies for recovering stellar mass.} There are only minor differences in stellar mass and SFR in this case as those are results that these galaxy models are tuned on, whereas the assumptions for dust mass cause in the distributions in the top right panel to be different.}
    \label{fig:simba_train_combined}
\end{figure*}

We next explore the generalization of \modelname to novel data sets to investigate the epistemic uncertainty in our model. Epistemic uncertainty refers to systematic uncertainty tied to modeling limitations which cannot be reduced by adding more training data (e.g. stellar isochrones). We first prepare alternative ML models trained on only \textsc{simba} galaxy photometry and tested on \textsc{IllustrisTNG} galaxy photometry to begin to evaluate the extent to which the output results from \modelname depend on the availability of modeling differences in training set. In Figure~\ref{fig:simba_train_combined}, we compare the performance of these models trained only on \textsc{simba} galaxies spectra and evaluated on \textsc{IllustrisTNG} galaxies to the performance of the fiducial model on the fiducial training sets. We find that in these tests there does exist a bias from having a limited training set associated with only one galaxy simulation model, even if we marginalize over the range of galaxy physics uncertainties in the CAMELS 1P model. The scale of these biases can largely be traced back to the inherent differences in the physical properties represented in the physics models we present in in Figure~\ref{fig:sim_sum_grid}; the distributions of stellar mass and SFR between the fiducial \textsc{simba} and \textsc{IllustrisTNG} are similar because they are tuned to reproduce observations, so the bias from only having one galaxy formation model available for training is minimal for these properties by construction. However, because of the differences in the dust masses that come from forward modeling the two galaxy formation model results we discuss in Section \ref{subsubsec:simrad} (\textsc{IllustrisTNG} galaxies typically have more dust assumed at the same stellar mass), the \textsc{simba}-trained model is biased towards underestimating the dust mass in the \textsc{IllustrisTNG} testing set relative to the fiducial model trained on the dust mass distribution from both simulation suites. This result represents a real concern about whether a simulation-based SED modeling tool, even trained from self-consistent galaxy formation models, can generalize to real galaxies given the many assumptions and subresolution models.

\begin{figure*}
     \plotone{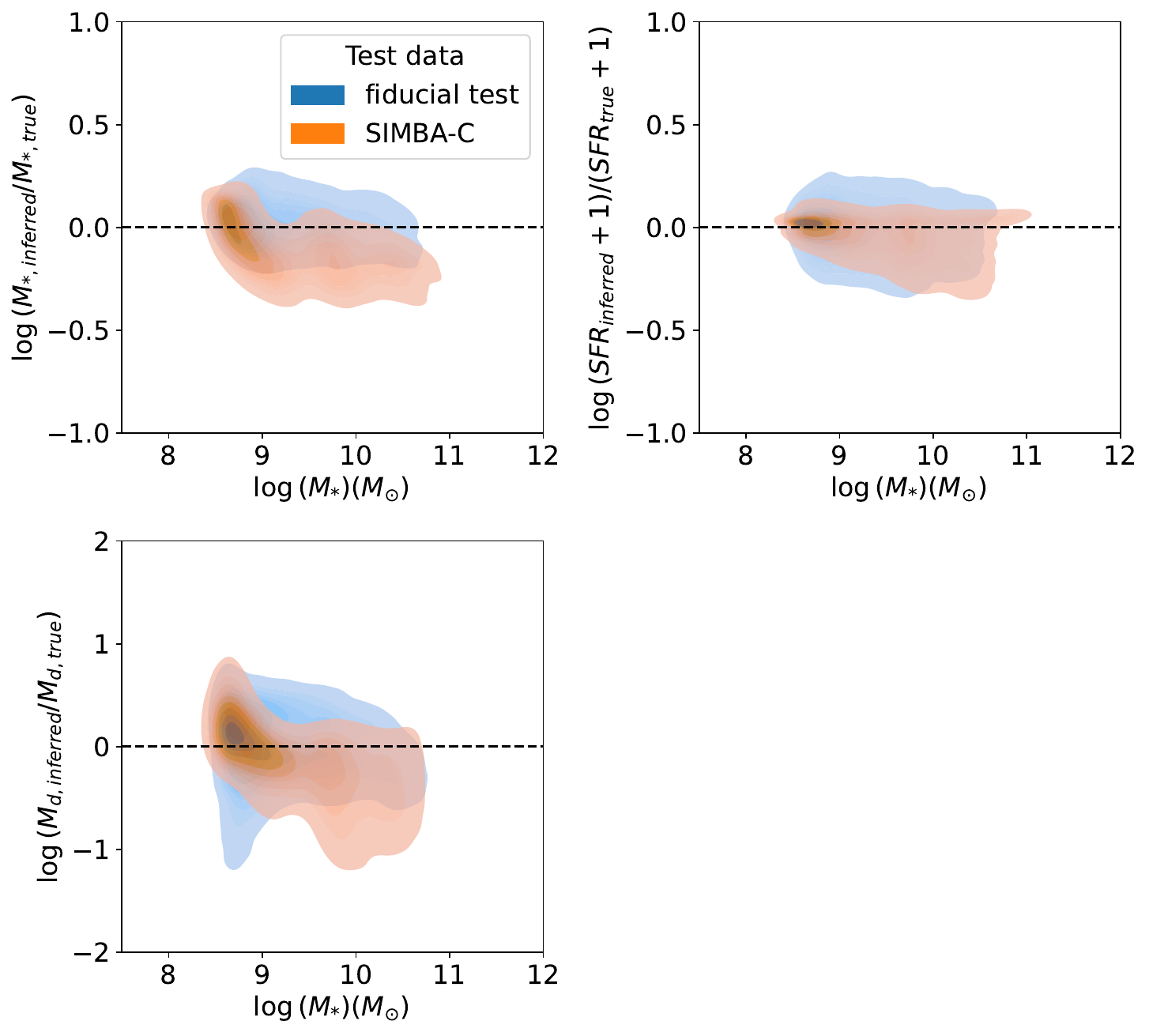}[t]
     \caption{\textbf{\modelname shows mixed results in recovering the ground truth of a dataset constructed on the spectra of \textsc{Simba-C} galaxies.} Top left: 2D KDE plot of accuracy in recovering ground truth stellar mass as a function of stellar mass. Top right: 2D KDE plot of accuracy in recovering ground truth SFR as a function of stellar mass. Bottom left: 2D KDE plot of accuracy in recovering ground truth dust mass as a function of stellar mass.}
     \label{fig:simbac_train_combined}
\end{figure*}

We also perform a generalization test using the \textsc{simba-c} cosmological simulation \citep{hough_simba-c_2023}. \textsc{simba-c} represents an update to the physics of \textsc{simba}, mainly in the new stellar feedback and chemical enrichment model Chem5 now implemented in \textsc{Gizmo} \citep{kobayashi_simulations_2007,kobayashi_origin_2020,kobayashi_new_2020}. Additionally, the update also changes the wind velocity scalings, and black hole seeding strategy. We treat the spectra generated from these galaxies as a novel test set, given that this model is not represented in the training set and the updates to all the stellar processes represent major changes in the galaxy physics models as compared to \textsc{simba}, though note that the dust model has not been changed or tuned in \textsc{simba-c}. Notably, the \textsc{simba-c} box is 100 Mpc/h on a side, which allows a greater number of galaxies with larger stellar masses to grow that would not form from fiducial galaxy physics within our training set derived from CAMELS boxes at 25 Mpc/h on a side. We forward model the $z=0$ galaxy spectra with \textsc{powderday} and redshift them to $z=1$ to because of the limitations in the training set discussed in Section~\ref{subsec:prosp}. We then evaluate the performance of \modelname on these sets of photometry.

In Figure~\ref{fig:simbac_train_combined}, we compare the performance on the novel \textsc{simba-c} data of \modelname to its performance on the test set for our `perfect information' test case. \modelname typically underestimates the stellar mass of the \textsc{simba-c} galaxies; we note that despite differences in the stellar mass ranges of the two datasets, \modelname still struggles to perform well on stellar masses well-represented in the training set. \modelname recovers \textsc{simba-c} dust masses more successfully than it even does on the fiducial test set, though it has broader tails than the distribution of the performances on the fiducial test set. The likely reason it lacks the bimodal feature is simply because there are no galaxies from \textsc{IllustrisTNG} in the \textsc{simba-c} set, which have a different relation between their stellar mass and dust mass. For less massive galaxies, \modelname tends to slightly overpredict the dust mass, while for more massive galaxies \modelname is slightly biased to underpredict the mass. There is a subset of the most massive galaxies in the \textsc{simba-c} sample that \modelname struggles with more greatly; these are likely the most massive quenched galaxies from \textsc{simba-c}, a population not well-represented in the training set because of the limited box size. \modelname also reproduces SFR well; there is a minor bias which seems to align with the peak of the distribution within the training set. Overall, these figures suggest that there is mixed performance with the generalization of \modelname to novel data, especially for galaxies whose properties are not well-represented in the data.

\section{model caveats and limitations} 
We now outline some of the important and limitations caveats in the construction of \modelnamec. First, we discuss potential concerns regarding the the training set. \modelname is trained on a subset of the CAMELS 1P 25 Mpc/h boxes. For computational feasibility, the resolution was reduced in these runs relative to the flagship \textsc{simba} and \textsc{IllustrisTNG} runs at similar box sizes. A combination of small boxes at low resolution sets the resolution limit of galaxies at $\sim10^{8.5} M_{\odot}$ and causes galaxies above $\sim10^{10.5} M_{\odot}$ to be rare in the sample. Additionally, any galaxies that do have masses greater than this range are likely tied to the CAMELS 1P volumes with the most extreme reductions in stellar or AGN feedback. Unfortunately, an input galaxy whose mass lies outside this $\sim$2 dex range might instead be assigned a redshift and mass that is incorrect as a consequence. Some CAMELS 1P runs may also produce galaxy populations at $z=0$ that are not necessarily reflective of what one would expect in the local universe, so they may introduce some is some potential bias into the training set in our attempt to marginalize over uncertainties in galaxy formation physics. For example, varying the black hole seed mass in \textsc{simba} results in few massive galaxies at $z=0$ (though the same is not true for the same change in \textsc{IllustrisTNG}). Future versions of \modelname could make a more deliberate effort to remove galaxies whose formation and evolution histories significantly differ than what we expect based off of observations. Overall, \modelname being trained a subset of the CAMELS 1P set makes it ill-suited for modeling both massive and dwarf galaxies or galaxies with rare formation histories.

The simulation sample limitations could be also exacerbated in the imputation step of \modelname because of the available galaxy populations in the CAMELS boxes. Despite the fact that we impose a limit on the number of reference spectra for fixed redshift in the KNN imputer model so that there is not a bias towards there being more low-redshift galaxies, there are still naturally a higher fraction of lower mass galaxies in our models. In a case where one has limited photometry and no knowledge of the redshift, a massive galaxy may have more `neighbors' in feature space at lower masses and redshifts, and we may not capture the full diversity of massive galaxy spectra in our more limited sample.  Additionally, the galaxies in our training set at non-integer redshifts are generated from integrating the photometry from integer redshifts at those intermediate redshifts. This decision means at non-integer redshifts, the set of available spectra was not produced from self-consistent star formation histories.

Next, we consider limitations with respect to our radiative transfer. We have naturally assumed a set of stellar models, an IMF, and a dust extinction law to run radiative transfer (see \ref{subsubsec:simrad}). However, stellar population synthesis modeling has known systematic uncertainties associated with it, and it is likely that if we had chosen a different stellar library, our inference of galaxy parameters would be systematically offset. The IMF is critically important to translating individual stellar spectra to the spectrum of a stellar population, and dust extinction requires assuming a set of dust composition and grain size distribution. Though these assumptions certainly impact our results and it is likely that they do not apply for all galaxies at all redshifts in our Universe, there is no current agreed-upon strategy to avoid these assumptions in either our modeling strategy or traditional SED fitting. Furthermore, although our radiative transfer does include star-dust geometry effects in the spectra, the limited resolution of the simulation boxes means that the star-dust geometry may not be as complex as in real galaxies. Finally, we do not model galaxy AGN emission or emission lines from HII regions in our spectra. Therefore, \modelname will not be appropriate to apply to galaxies that have significant emission from their SMBHs. We do not model emission lines because they are highly dependent on uncertain subresolution assumptions, and because they come at significant computational cost.

We finally shift from discussing the limitations that originate in our training set to the limitations in related to the structure and components of \modelnamec. As discussed in Section~\ref{subsec:model}, while KNN imputation is generally successful in \modelname to deal with missing input filters, the KNN imputer does not naturally produce uncertainties. We have attempted to sample within the photometric errors associated with an observation so that we could explore the space of potential similar photometric shapes within the training set, but our error analysis suggests that the majority of the output uncertainty is attached to the distribution produced by the {\sc NGBoost} algorithm and the additional uncertainty this imputation sampling strategy adds does not fully reflect the loss in accuracy (see Section~\ref{subsec:ci_invest}). Future versions of \modelname would ideally include an imputation strategy that returns realistic photometry along with uncertainties as an output. Additionally, as the number of inputs scales AND as the amount of training data scales, the KNN imputer becomes a speed bottleneck.  A well-designed and trained de-noising autoencoder (DAE) could address the uncertainty concern in addition to the KNN speed bottleneck. 

Furthermore, {\sc NGBoost} lacks flexibility in several key aspects. {\sc NGBoost} is only capable of producing outputs Gaussian distributions as predictions. Other, more sophisticated, ML-based algorithms based on neural networks and designed to predict uncertainties, such normalizing flows, can output complex PDFs that reflect the complex multivariate modeling of spectra. Based on the results of our SHAP analysis (see Section~\ref{subsec:shap}), the {\sc NGBoost} models within \modelname heavily rely on certain photometric features to infer galaxy properties. Though we identify physical justifications for this behavior, it results in the predictions of \modelname being highly dependent on the success of the imputer in these bands. Future versions of our model could instead use a neural network with random dropout in training so that the model is better equipped to infer results from a variety of photometric information instead of one filter in particular.

\section{Summary}\label{sec:sum}

 We conclude by summarizing our main results.

\begin{itemize}
  \setlength\itemsep{1em}
    \item We present our SBI-based SED modeling package \modelname designed to infer three galaxy physical properties - stellar mass, dust mass, and star formation rate. The ML models within \modelname are trained on radiative transfer run on a subset of the CAMELS 1P cosmological simulation suite galaxies for integer redshifts $z=0-6$. Importantly, \modelname returns uncertainties in its outputs and is able to handle variable amounts of input photometry by using a KNN imputation strategy.
    \item We find that \modelname outperforms traditional SED fitting on the test set by every evaluation metric we use for each physical property and for each photometric test we run (Section~\ref{sec:res},Table~\ref{table:all_metric}).
    \item \modelname has inferred confidence intervals for galaxy properties that contain the ground truth more often than traditional SED fitting in the cases we test. However, in the cases where only limited photometry is available, the quoted uncertainties less often mirror the offset from the ground truth (Section~\ref{subsec:jwst_phot}, Appendix~\ref{sec:HST_phot}). The ML models' predicted probability distributions are the primary sources of the size of the output uncertainties (Section~\ref{subsec:ci_invest}, Figure~\ref{fig:unc_break}), but sampling within input photometry uncertainties does cause the size of the CIs to grow as the amount of constraining information decreases. However, this growth does not scale sufficiently with the loss in accuracy.
    \item We perform a SHAP analysis to understand the photometric points which the ML models within \modelname are relying on to make inferences about galaxy properties and find that the results largely follow natural physical explanations (Section~\ref{subsec:shap}, Figures~\ref{fig:shap_sm}, \ref{fig:shap_dm}, and \ref{fig:shap_sfr}).
    \item We investigate the galaxies that the models perform the worst on and find that they typically are (1) on most/least massive extremes of the galaxies in the training set and/or (2) an outlier in the distribution of stellar ages relative to the training set (Section~\ref{subsec:bad_gals}, Figure~\ref{fig:outlier}).
    \item We investigate the generalization of the model (Section~\ref{subsec:train_test}). We first train an additional set of models only on \textsc{simba} spectra and evaluate their performance on \textsc{IllustrisTNG} spectra. The most notable bias in this test is in the dust mass, which comes from the difference in the stellar mass - dust mass relation between the two simulations (Figure~\ref{fig:simba_train_combined}). We then evaluate the fiducial model on the \textsc{simba-c} cosmological simulation and find that while \modelname is successful in recovering the SFR and dust mass, it struggles to recover the stellar mass of \textsc{simba-c} galaxies (Figure~\ref{fig:simbac_train_combined}). Both of these results raise concerns about whether \modelname can successfully generalize to novel data. 
\end{itemize}

\section{Data Availability}
The galaxy properties and photometry from the CAMELS 1P suite used to train the models with \modelname are available upon reasonable request. \modelname can be installed and downloaded from \url{https://github.com/DhruvZ/Phot-Gal}.

\section{Acknowledgements}
The authors acknowledge UFIT Research Computing for providing computational resources on HiPerGator and support that have contributed to the research results reported in this publication (\url{http://it.ufl.edu/rc}). D.Z. was funded by a grant from the UF Astraeus Institute. D.N. acknowledges funding from NASA via grants ATP-21-0013 and ATP-23-0002. D.N. also acknowledges funding from grant HST-AR-16626. D.N. additionally thanks the Aspen Center for Physics which is supported by National Science Foundation grant PHY-1607611, which is where the original framework for the {\sc powderday} code base was developed. Finally, D.N. is grateful to Charlie Conroy, Ben Johnson and Joel Leja for their guidance over the years in discussions about galaxy property inference.

\bibliography{references}

\appendix

\section{HST photometry/unknown redshift}\label{sec:HST_phot}

\begin{figure*}
    \epsscale{0.75}
    \plotone{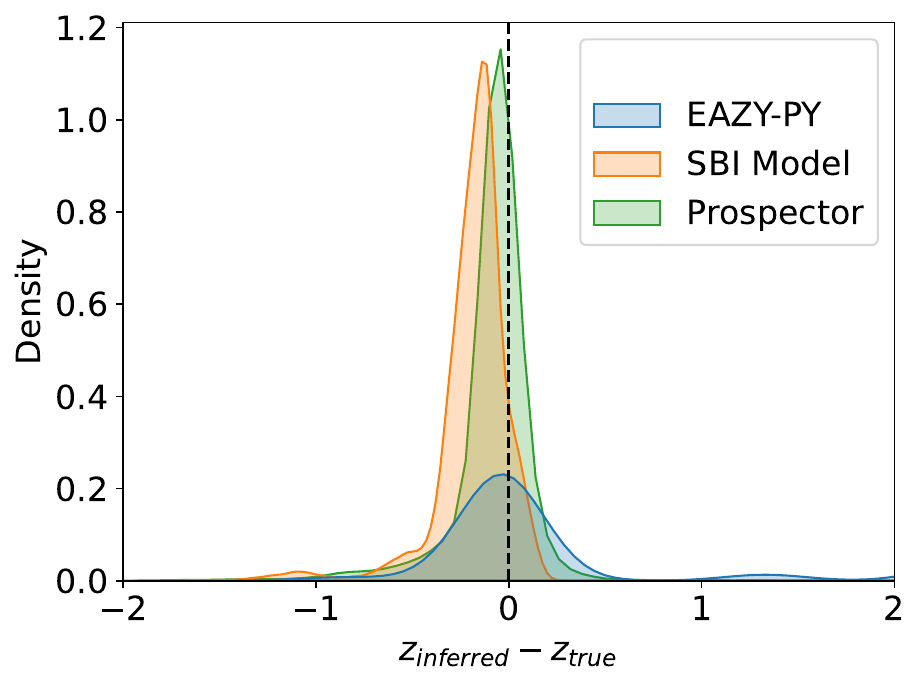}
    \caption{\textbf{Comparison of the performance of the \modelname with \textsc{eazy-py} and \textsc{Prospector} for recovering the true redshift with HST photometry.} }
    \label{fig:z_HST}
\end{figure*}

In this appendix, we compare the performance with only HST photometry available. Specifically, we limit the input photometry to the Hubble Wide Field Camera 3 (WFC3) photometric filters, again with an assumed SNR of 10 and treat the redshift as unknown.
In Figure~\ref{fig:z_HST}, we note that \modelname is once again reasonably reproducing the true redshift with photometric data, though it typically is biased to slightly underestimate the true redshift. In comparison with the case where JWST photometry was available, \textsc{eazy-py} performs significantly worse in recovering the true redshift. However, \textsc{prospector} performs better than \modelname and \textsc{eazy-py} in recovering the true redshift.

\begin{figure*}
    \epsscale{1.25}
    \plotone{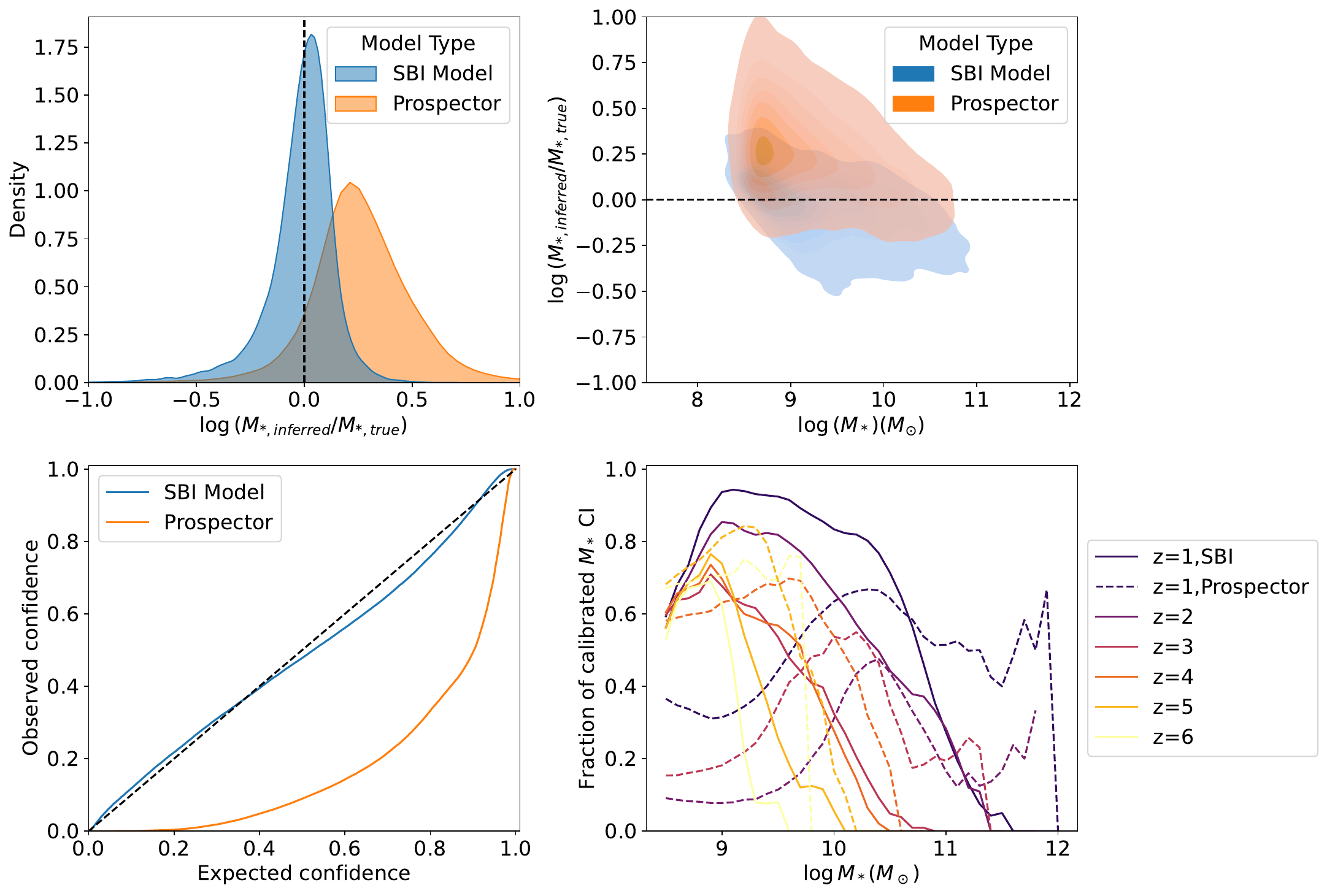}
    \caption{\textbf{Similar to Figure~\ref{fig:mstar_jwst}, but now with only HST photometry available and the redshift treat as unknown for model inputs.} }
    \label{fig:mstar_HST}
\end{figure*}

With only HST photometry, \modelname in Figure~\ref{fig:mstar_HST} is slightly biased to overpredict stellar masses, but overall the performance is still reasonable. Additionally, the bias where the lowest mass galaxies are overpredicted and the highest mass galaxies are underpredicted has worsened. In the limiting ideal case in Section~\ref{subsec:full_phot}, we noted that there is a characteristic stellar mass for each redshift where the fraction of calibrated CIs rapidly falls off. In this example case with only HST photometry available, that characteristic mass at which the reported uncertainties become untrustworthy is much lower. In theory, the lack of mid-IR to far-IR constraints should add significantly uncertainty to the imputed IR and consequently broaden the posterior (also seen in Figure~\ref{fig:knn_imp}). However, reported CIs do not grow sufficiently with the loss of information. This implies that the imputation step is not adequately adding to the posterior (see Section~\ref{subsec:ci_invest} as well for discussion of the uncertainty components).\textsc{prospector} exhibits a typical bias of $\sim0.25$ dex in recovering the true stellar mass, but a higher fraction of the stated uncertainties are appropriate for the bias relative to the complete information case. Unlike that case, where the fraction of CIs at different redshifts and stellar masses that reflect the true uncertainty holds relatively constant, the lower redshift galaxies in the test set tend to have a higher fraction of correctly sized uncertainties.

\begin{figure*}
    \epsscale{1.25}
    \plotone{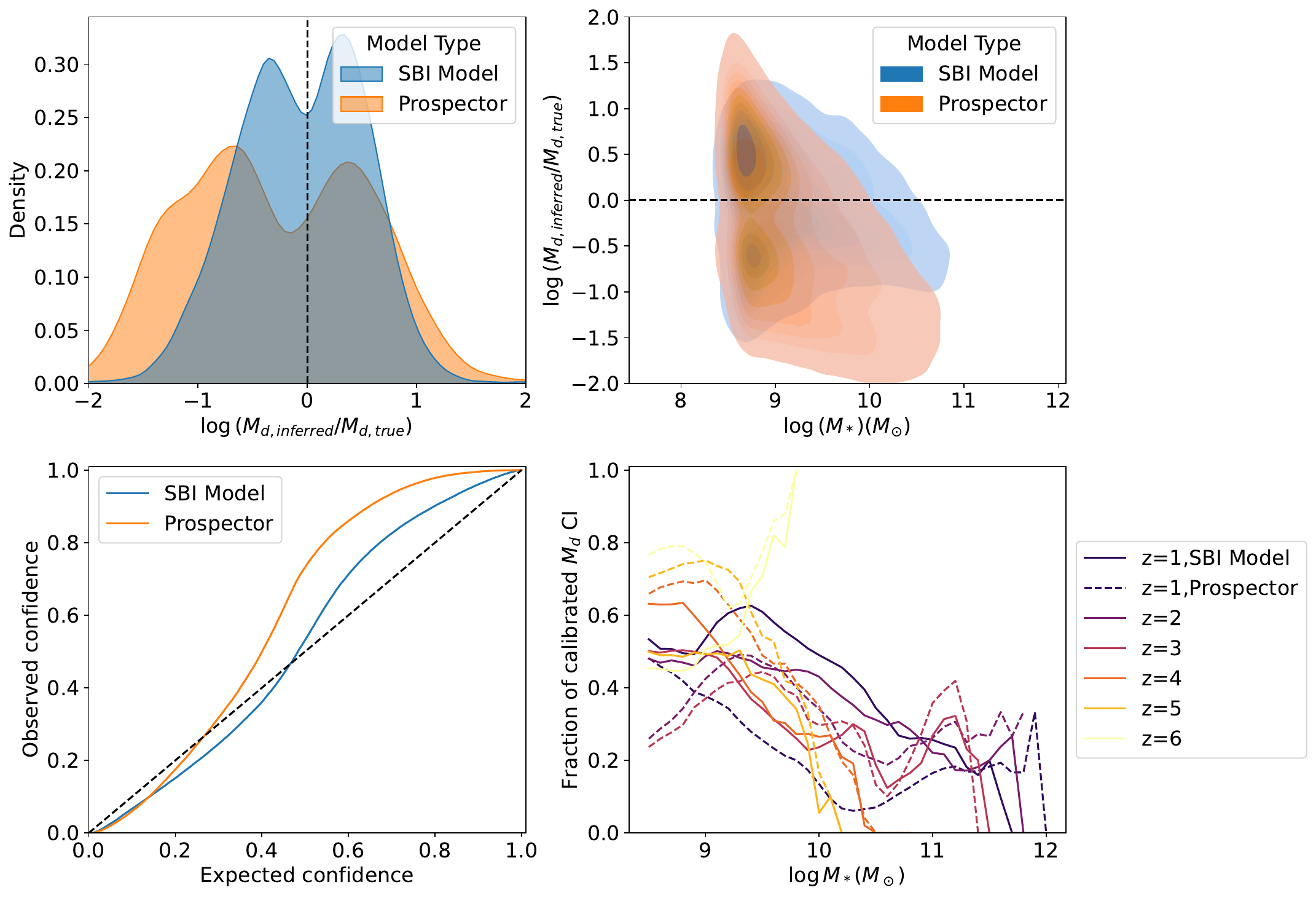}
    \caption{\textbf{Similar to Figure~\ref{fig:mstar_HST}, but for galaxy dust mass inference with HST photometry only and no redshift information as model inputs.}}
    \label{fig:md_HST}
\end{figure*}

Unsurprisingly, without constraints on the IR, the dust mass accuracy of \modelname in Figure~\ref{fig:md_HST} for the HST photometry only case is the worst up to this point, with some galaxies up to $\sim 2$ dex offset from the true dust mass. \textsc{prospector} performs qualitatively similarly with a slightly broader distribution. Furthermore, the fraction of reasonable stated uncertainties noticeably decreases above $10^{10} M_{\odot}$. Neither \modelname nor \textsc{prospector} is likely to have the confidences intervals contain the ground truth, and their performance is relatively similar across redshift.

\begin{figure*}
    \epsscale{1.25}
    \plotone{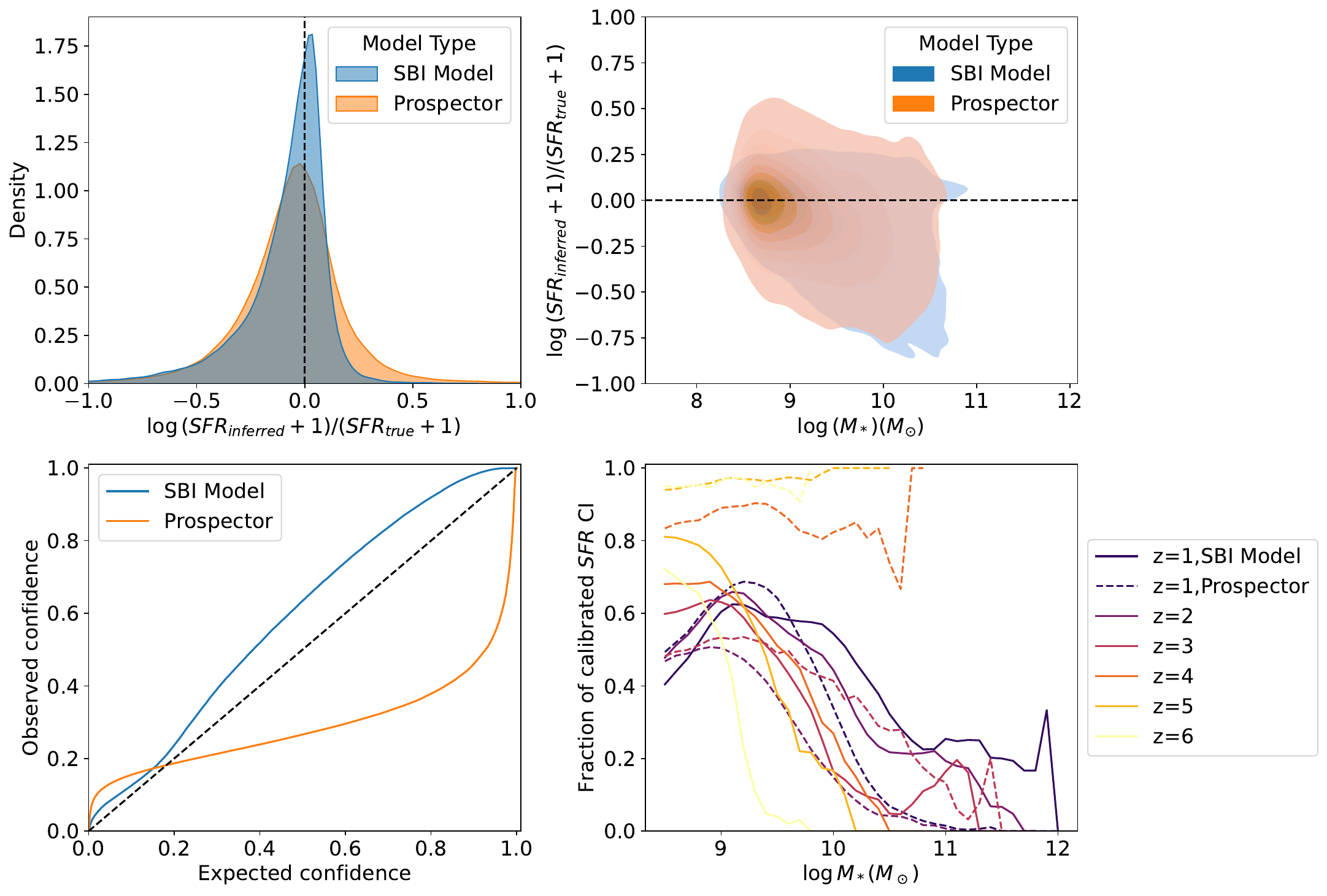}
    \caption{\textbf{Similar to Figure~\ref{fig:mstar_HST}, but for galaxy SFR inference with HST photometry only and no redshift information as model inputs.}}
    \label{fig:sfr_HST}
\end{figure*}

Without the IR constraints, both \modelname and \textsc{prospector} do a decent job of recovering the SFR (Figure~\ref{fig:sfr_HST}). Both are biased to underpredict the SFR of the most massive galaxies in the test set, but the distribution of performance for both is broader than for the best-case scenario. While the peak of the distribution is still centered on 0, there is a clear skew to the distribution, corresponding to poorer performance on predicting the SFR of higher mass systems with only HST photometry. \modelname more often recovers appropriate uncertainties for the lowest mass galaxies in the sample and the performance progressively decreases from there. \textsc{prospector} behaves similarly for $z=1-3$, but for $z=4-6$, the stated uncertainties are accurate $\gtrsim80\%$ of the time.

\end{document}